\documentclass[a4paper,11pt]{article}
\pdfoutput=1 % if your are submitting a pdflatex (i.e. if you have
\usepackage{jheppub} % for details on the use of the package, please
\usepackage[T1]{fontenc} % if needed
\usepackage{shuffle}
\usepackage{slashed}
\usepackage{tikz}
\usepackage{adjustbox}
\usepackage{array}

\usetikzlibrary{shapes,decorations.markings}

\newcommand{\bea}{\begin{eqnarray}}
\newcommand{\eea}{\end{eqnarray}}
\newcommand{\bean}{\begin{eqnarray*}}
\newcommand{\eean}{\end{eqnarray*}}
\newcommand{\nn}{\nonumber\\}

\def\Label#1{\label{#1}%
  \smash{\hbox to0pt{\raise1ex\hbox{\tiny[#1]}\hss}}}

\newcommand {\Pf}  {\text{Pf}\,}
\newcommand {\Pfp}  {\text{Pf}\,'}
\newcommand {\PT} {\text{PT}\,}

\title{BCJ numerators from reduced Pfaffian}

\author[a]{Yi-Jian Du}
\author[b]{anc Fei Teng}

\affiliation[a]{Center for Theoretical Physics, School of Physics and Technology, Wuhan University \\ No. 299 Bayi Road, Wuhan 430072, P. R. China}
\affiliation[b]{Department of Physics and Astronomy, University of Utah \\ 115 South 1400 East, Salt Lake City, UT 84112, USA}

\emailAdd{yijian.du@whu.edu.cn}
\emailAdd{Fei.Teng@uath.edu}

\abstract{By expanding the reduced Pfaffian in the tree level Cachazo-He-Yuan (CHY) integrands for Yang-Mills (YM) and nonlinear sigma model (NLSM), we can get the Bern-Carrasco-Johansson (BCJ) numerators in Del Duca-Dixon-Maltoni (DDM) form for arbitrary number of particles in any spacetime dimensions. In this work, we give a set of very straightforward graphic rules based on spanning trees for a direct evaluation of the BCJ numerators for YM and NLSM. Such rules can be derived from the Laplace expansion of the corresponding reduced Pfaffian. For YM, the each one of the $(n-2)!$ DDM form BCJ numerators contains exactly $(n-1)!$ terms, corresponding to the increasing trees with respect to the color order. For NLSM, the number of nonzero numerators is at most $(n-2)!-(n-3)!$, less than those of several previous constructions.}

\keywords{Scattering Amplitude}

\begin{document}
\maketitle
\flushbottom

%%%%%%%%%%%%%%%%%%%%%%%%%%%%%%%%%%%%%%%%%%%%%%%%%%%
\section{Introduction}
%%%%%%%%%%%%%%%%%%%%%%%%%%%%%%%%%%%%%%%%%%%%%%%%%%%

As first pointed out by Bern, Carrasco and Johansson (BCJ)~\cite{Bern:2008qj}, the kinematic numerators in tree level Yang-Mills (YM) amplitudes can satisfy a secret algebra that enjoys the same Jacobi identity as Lie algebras. Once such numerators are found, a double copy of them directly gives the tree level Einstein gravity amplitudes~\cite{Bern:2010ue}. More recently, Cachazo, He and Yuan (CHY) proposed a new formalism for tree level amplitudes of a variety of theories~\cite{Cachazo:2013gna,Cachazo:2013hca,Cachazo:2013iea,Cachazo:2014nsa,Cachazo:2014xea}. As a general feature of the CHY formalism, the kinematic and polarization information of a given theory are usually packed into a reduced Pfaffian, while the color or flavor ordering is captured by a Parke-Taylor factor (see Section~\ref{sec:CHY} for details). The CHY formalism makes manifest the double copy relations between gauge and gravity theories. It also points out a way to obtain directly the BCJ numerators~\cite{Cachazo:2013iea} (also in~\cite{Naculich:2014rta} from a different perspective): expand the reduced Pfaffian by the Kleiss-Kuijf (KK) basis~\cite{Kleiss:1988ne} and the coefficients are just what we want.\footnote{To be specific, what we get are the BCJ numerators in Del Duca-Dixon-Maltoni (DDM) form~\cite{DelDuca:1999rs}, see Section~\ref{sec:BCJ}.} However, it is hard to find a well-controlled way to write down the final result of the expansion for arbitrary number of particles. The construction of BCJ numerators has been studied from the kinematic algebra~\cite{Monteiro:2013rya} and the reduction of CHY integrals~\cite{Bjerrum-Bohr:2016juj,Bjerrum-Bohr:2016axv}.

In this work, we are going to derive the BCJ numerators from a systematic expansion of the reduced Pfaffians. There are in principle two ways to disentangle the Pfaffian of a $2n\times 2n$ antisymmetric matrix $X=(x_{ij})$: using the formal definition
\begin{equation}
\label{eq:formal}
    \Pf(X)=\frac{1}{2^nn!}\sum_{\pmb\sigma\in S_{2n}}\text{sign}(\pmb\sigma)\prod_{i=1}^{n}x_{\sigma(2i-1)\sigma(2i)}\,,
\end{equation}
or the Laplace expansion
\begin{equation}
    \Pf(X)=\sum_{j=1,j\neq i}^{2n}(-1)^{i+j+1+\theta(i-j)}x_{ij}\Pf(X_{ij}^{ij})\,,\qquad\theta(i-j)\text{ is a step function,}
\end{equation}
where $X_{ij}^{ij}$ means that we have deleted the $i$-th and $j$-th row and column from $X$. Lam and Yao have used the formal definition \eqref{eq:formal} to fully expand the reduced Pfaffian by the basis corresponding to the cycles in the permutation group~\cite{Lam:2016tlk}.

On the other hand, the Laplace expansion of Pfaffian enables us to attack the problem from the recursive aspect. The general pattern we have obtained is the following. For the amplitudes with $n$ particles belonging to the adjoint of the gauge (or flavor) Lie algebra, the Laplace expansion will lead to a linear combination of the amplitudes with $s$ bi-adjoint scalars and $n-s$ original adjoint particles (with all possible $s$). The expansion coefficients contain the information of kinematics and polarization of the adjoint particles. The reason is that the original adjoint amplitude can be represented by the product of a reduced Pfaffian and a Parke-Taylor factor. The Laplace expansion recursively pulls out several entries from the Pfaffian, reducing the matrix size. Next, we strip off the kinematic and polarization from those entries that get pulled out, viewing them as coefficients, and group the rest into another Parke-Taylor factor. Now we have a combination of amplitudes represented by a product of two Parke-Taylor factors and a smaller Pfaffian. Essentially, they are the amplitudes of some bi-adjoint scalars interacting with the adjoint particles. If we further carry out this recursion to each of the amplitudes in the resultant linear combination, at the end we will get a linear combination of pure bi-adjoint amplitudes, whose coefficients are just the BCJ numerators we want. The advantage of this approach is that by using this step-by-step recursion, we can easily summarize a set of rules for constructing the final BCJ numerators directly. Such rules would be obscured without the help of the recursive relation. In particular, these rules are difficult to extract in Lam and Yao's expansion~\cite{Lam:2016tlk}.

By using the Laplace expansion, Feng and one of the current authors~\cite{Teng:2017tbo} have successfully expand the single trace Einstein-Yang-Mills (EYM) amplitudes in terms of the KK basis pure YM amplitudes (or equivalently, YM-scalar amplitudes in terms of bi-adjoint scalar ones). The coefficients can be very nicely evaluated from a set of graphic rules based on spanning trees. Then in~\cite{Fu:2017uzt}, a very simple linear relation between YM amplitudes and YM-scalar amplitudes is derived, which enables us to write down a similar set of graphic rules to directly evaluate the BCJ numerators for YM. This construction will give polynomial form BCJ numerators in the DDM form, which will be studied in detail in Section~\ref{sec:rules}. Interestingly, each numerator contains exactly $(n-1)!$ terms for $n$ particles, corresponding to a special set of $n$-point spanning trees.

Then we show in Section~\ref{sec:NLSM} that the same technique also applies to NLSM. Actually, the same set of graphic rules for the BCJ numerators of NLSM can be obtained through a dimensional reduction from those YM rules~\cite{Huang:2017ydz}.\footnote{We note that this scheme is a little different from the original one proposed in~\cite{Cachazo:2014xea}.} Following these rules, we always get less than $(n-2)!-(n-3)!$ nonzero numerators for $n$-point NLSM. Each numerator contains a sum over a subset of the corresponding YM spanning trees. The essential graph theoretical properties of these trees are also discussed.

The structure of this paper is as follows. Section~\ref{sec:BCJ} reviews the color-kinematic duality and double copy construction. We then discuss the CHY integrands involved in this work in Section~\ref{sec:CHY}. Then in Section~\ref{sec:rules} and \ref{sec:NLSM}, we discuss the recursive expansion of the reduced Pfaffians, and the graphic rules for constructing BCJ numerators for YM and NLSM respectively. Some useful details in our calculation are put in Appendix~\ref{sec:laplace} and \ref{sec:derivation}. Finally, the full explicit results of $5$-point YM numerators are shown in Appendix~\ref{sec:numerator}, and $6$-point NLSM numerators in Appendix~\ref{sec:NLSMnum}.

%%%%%%%%%%%%%%%%%%%%%%%%%%%%%%%%%%%%%%%%%%%%%%%%%%%
\section{Color-kinematic duality and double copy construction}\label{sec:BCJ}
%%%%%%%%%%%%%%%%%%%%%%%%%%%%%%%%%%%%%%%%%%%%%%%%%%%

Tree level total YM scattering amplitudes $\mathcal{A}_{n}$ can be formally expressed as
\begin{equation}
    \mathcal{A}_{n}=\sum_{i\in\text{cubic}}\frac{c_i n_i}{D_i}\,,
\end{equation}
where the sum is over all cubic trees with $n$ external legs, which can be identified as the usual Feynman diagrams.\footnote{Feynman diagrams in general contain quartic vertices, but we can blow them up into two cubic vertices by multiplying and dividing appropriate propagators.} In this equation, $1/D_i$ is the product of propagators associated with each diagram, while $c_i$ and $n_i$ are respectively the color factor and kinematic numerator. From Feynman rules, one can tell that $c_i$ is a chain of structure constants $f^{abc}$, determined by the gauge group, while $n_i$ is composed of the scalar products of external momenta and polarization vectors. The numerator $n_i$ is not unique, in the sense that a generalized gauge transformation $n_i\rightarrow n_i+\Delta_i$ is allowed if:
\begin{equation}
    \sum_{i\in\text{cubic}}\frac{c_i\Delta_i}{D_i}=0\,,
\end{equation}
which holds by using only the color Jacobi identity of $c_i$. We note that $\Delta_i$ may result from a usual gauge transformation by replacing $\epsilon$ by $k$ in $n_i$, or a very nontrivial field redefinition at the Lagrangian level.

Bern, Carrasco and Johansson showed in~\cite{Bern:2008qj} that the above mentioned generalized gauge freedom allows us to write down a set of numerators (BCJ numerators) that satisfy the same Jacobi identity as the color factors, without changing the total amplitude $\mathcal{A}_{n}$:
\begin{align}
\label{eq:BCJnumerator}
    n_i=-n_j\;\Longleftrightarrow\; c_i=-c_j\,,& &n_i+n_j+n_k=0\;\Longleftrightarrow\; c_i+c_j+c_k=0\,,
\end{align}
namely, there exists a color-kinematic duality. The existence of such numerators are guaranteed by the BCJ relation of the color ordered amplitudes. Moreover, once these BCJ numerators are found, we can simply replace the color factors by them and obtain the Einstein gravity amplitudes~\cite{Bern:2010ue}:
\begin{equation}
\label{eq:double}
    M_{n}=\sum_{i\in\text{cubic}}\frac{n_i\widetilde{n}_i}{D_i}\,.
\end{equation}
In this expression, we only require one of the numerators, say $n_i$, to satisfy the duality, while $\widetilde{n}_i$ can be any set of valid YM numerators. More generally, once we have two gauge theories (with matter interactions or supersymmetry extension) that both can satisfy the color-kinematic duality, double copy constructions will always lead to gravity scattering amplitudes~\cite{Johansson:2015oia,Chiodaroli:2014xia,Cachazo:2014xea,Chiodaroli:2015rdg,Chiodaroli:2015wal,Cheung:2016prv,Chiodaroli:2017ngp}.

The construction of BCJ numerators starts from the following observation. Eq.~\eqref{eq:BCJnumerator} indicates that there are linear relations among the numerators in the space of cubic graphs, due to the Jacobi identity. For the color factor $c_i$, Del Duca, Dixon and Maltoni (DDM)~\cite{DelDuca:1999rs} showed that the independent basis under the Jacobi identity is the set of half-ladder diagrams, with leg $1$ and $n$ fixed. The total amplitude in this basis has the expansion
\begin{equation}
    \mathcal{A}_{n}=\sum_{\pmb\sigma\in S_{\{2\ldots n-1\}}}f^{a_1a_{\sigma(2)}b_1}f^{b_1a_{\sigma(3)}b_2}\ldots f^{b_{n-3}a_{\sigma(n-1}a_n}A_{n}(1,\pmb\sigma,n)\,,
\end{equation}
where $A_n$ is the color ordered YM amplitude, and $b_i$'s are dummy indices got summed over implicitly. Since deriving this result only involves the Jacobi identity, we can perform the same manipulation to Eq.~\eqref{eq:double} and write $M_n$ under the DDM form:
\begin{equation}
\label{eq:doublecopy}
    M_{n}=\sum_{\pmb\beta\in S_{\{2\ldots n-2\}}}n(1,\pmb\beta,n)A_{n}(1,\pmb\beta,n)\,.
\end{equation}
In other words, once we successfully expand the gravity amplitude in terms of the color ordered YM amplitudes in the KK basis~\cite{Kleiss:1988ne}, the expansion coefficients are just the BCJ numerators $n(1,\pmb\beta,n)$ associated to half-ladder diagrams (we will call them DDM form BCJ numerators in the following). Then we can work out all the BCJ numerators associated to generic cubic diagrams by repeatedly using of Jacobi identity.

The above story all applies to NLSM, in which $c_i$ is associated with a global flavor algebra. The double copy \eqref{eq:doublecopy} of NLSM leads to the special Galileon theory, as first indicated by the CHY formalism~\cite{Cachazo:2014xea}. The NLSM is expected to be simpler than YM since it is a scalar field theory with only a global flavor symmetry. Moreover, recent work by Arkani-Hamed, Rodina and Trnka~\cite{Arkani-Hamed:2016rak} showed that the soft limit behavior constrains the NLSM tree level amplitude in the same way as how the gauge invariance constrains the YM tree level amplitude. This helps one to understand why YM and NLSM shares some similar properties, for example, the color (flavor) kinematic duality and BCJ relations.

As the main subject of this work, we propose a systematic method for the direct evaluation of the DDM form BCJ numerators for both YM and NLSM, derived from the CHY formalism~\cite{Cachazo:2013gna,Cachazo:2013hca,Cachazo:2013iea,Cachazo:2014nsa,Cachazo:2014xea}. Before that, we need to introduce the CHY representation for YM, gravity and EYM in the next section.

%%%%%%%%%%%%%%%%%%%%%%%%%%%%%%%%%%%%%%%%%%%%%%%%%%%
\section{CHY integrands and their relations}\label{sec:CHY}
%%%%%%%%%%%%%%%%%%%%%%%%%%%%%%%%%%%%%%%%%%%%%%%%%%%

In this section, we discuss the CHY integrands for $n$-point tree level YM and gravity amplitudes, as well as the single trace EYM amplitudes~\cite{Cachazo:2013gna,Cachazo:2013hca,Cachazo:2013iea,Cachazo:2014nsa,Cachazo:2014xea}. We would like to demonstrate that we can express them in terms of each other, and these relations lead to a very natural way to write down the BCJ numerator for YM in the DDM form.

\subsection{Definitions}
The central object in these CHY integrands is the $2n\times 2n$ matrix $\Psi$, defined as:
\begin{equation}
    \Psi=\left(\begin{array}{cc}
        A & -C^{T} \\
        C & B \\
    \end{array}\right)\,.
\end{equation}
The three $n\times n$ matrices $A$, $B$ and $C$ contained in $\Psi$ have the following forms:
\begin{align}
& A_{ab}=\left\{\begin{array}{>{\displaystyle}c @{\hspace{1em}} >{\displaystyle}l}
\frac{k_a\cdot k_b}{\sigma_{ab}} & a\neq b\\
0 & a=b \\
\end{array}\right.&
& B_{ab}=\left\{\begin{array}{>{\displaystyle}c @{\hspace{1em}} >{\displaystyle}l}
\frac{\epsilon_{a}\cdot\epsilon_{b}}{\sigma_{ab}} & a\neq b\\
0 & a=b \\
\end{array}\right.&
& C_{ab}=\left\{\begin{array}{>{\displaystyle}l @{\hspace{1em}} >{\displaystyle}l}
\frac{\epsilon_{a}\cdot k_{b}}{\sigma_{ab}} & a\neq b\\
-\sum_{c\neq a}\frac{\epsilon_{a}\cdot k_{c}}{\sigma_{ac}} & a=b \\
\end{array}\right.\,,
\label{eq:ABC}
\end{align}
where $1\leqslant a,b,c\leqslant n$. It is easy to show that $\Psi$ is antisymmetric and has co-rank two in its first $n$ rows and columns. We can thus delete two rows and columns from them and define the \emph{reduced Pfaffian} of $\Psi$ as:
\begin{equation}
\label{eq:Pfp}
    \Pfp(\Psi)=\frac{(-1)^{i+j}}{\sigma_{ij}}\Pf(\Psi_{ij}^{ij})\qquad 1\leqslant i<j\leqslant n+m\,.
\end{equation}
The matrix $\Psi_{ij}^{ij}$ is reduced from $\Psi$ by deleting the $i$-th and $j$-th  row and column. It can be proved that the value of $\Pfp(\Psi)$ does not depend on which two rows and columns are deleted~\cite{Cachazo:2013gna}. For EYM, there is another matrix involved:
\begin{equation}
\label{eq:EYMpf}
    \Psi_{\mathsf{H}}=\left(\begin{array}{cc}
        A_{\mathsf{H}} & -(C_{\mathsf{H}})^{T} \\
        C_{\mathsf{H}} & B_{\mathsf{H}} \\
    \end{array}\right)\,,
\end{equation}
where $\mathsf{H}=\{h_1h_2\ldots h_m\}$ represents the set of gravitons. The matrix $A_{\mathsf{H}}$, $B_{\mathsf{H}}$ and $C_{\mathsf{H}}$ are submatrices of $A$, $B$ and $C$ whose row and column indices take value only in $\mathsf{H}$. Generic CHY integrands have the form:
\begin{equation}
    \mathcal{I}_{\text{CHY}}=\mathcal{I}_{L}\mathcal{I}_{R}\,,
\end{equation}
and those integrands to be used in this work are shown in Table~\ref{tab:integrand}. To get the amplitude, we need to integrate it over a measure $d\Omega_{\text{CHY}}$ that imposes the scattering equation:
\begin{equation}
    \sum_{j\neq i}\frac{k_i\cdot k_j}{\sigma_i-\sigma_j}=0\,.
\end{equation}
In principle, we need $\Psi$ for theories involving polarization vectors; $A$ for adjoint scalar theories; and the Parke-Taylor factor:
\begin{equation}
    \PT(12\ldots n)=\frac{1}{\sigma_{12}\sigma_{23}\ldots\sigma_{n1}}\qquad\sigma_{ij}\equiv\sigma_{i}-\sigma_{j}\,.
\end{equation}
for color (flavor) orderings.

\begin{table}[t]
\centering
\renewcommand{\arraystretch}{1.1}
    \begin{tabular}{c|c|c|c}
         Theory & Integrand & $\mathcal{I}_L$ & $\mathcal{I}_{R}$ \\   \hline\hline
         Yang-Mills & $\mathcal{I}_{\text{YM}}$ & $\PT(12\ldots n)$ & $\Pfp(\Psi)$ \\
         Einstein gravity & $\mathcal{I}_{\text{GR}}$ & $\Pfp(\Psi)$ & $\Pfp(\Psi)$ \\
         Einstein-Yang-Mills (single trace) & $\mathcal{I}_{\text{EYM}}$ & $\PT(g_1g_2\ldots g_s)\Pf(\Psi_{\mathsf{H}})$ & $\Pfp(\Psi)$ \\
         Nonlinear sigma model & $\mathcal{I}_{\text{NLSM}}$ & $\PT(12\ldots n)$ & $\left[\Pfp(A)\right]^{2}$ \\
         Special Galileaon & $\mathcal{I}_{\text{Galileon}}$ & $\left[\Pfp(A)\right]^{2}$ & $\left[\Pfp(A)\right]^{2}$ \\
         $\text{NLSM}+\phi^{3}$ & $\mathcal{I}_{\text{NLSM}+\phi^{3}}$ & $\PT(12\ldots n)$ & $\PT(1,\pmb\alpha,n)\left[\Pf(A_{\overline{\pmb\alpha}})\right]^{2}$ \\
    \end{tabular}
    \label{tab:integrand}
    \caption{CHY integrands to be used in this paper. The notations will be explained when appear in the main text.}
\end{table}

A remarkable feature of the CHY formalism is that it makes the double copy construction very manifest, and provides a way to directly evaluate those BCJ numerators. As the first set of double copies, we consider the YM and Einstein gravity. Their tree level integrands are given by:
\begin{align}
    \mathcal{I}_{\text{YM}}(12\ldots n)=\PT\left(12\ldots n\right)\Pfp(\Psi)& &\mathcal{I}_{\text{GR}}(12\ldots n)=\Pfp(\Psi)\times\Pfp(\Psi)
\end{align}
As already proposed in~\cite{Cachazo:2013iea}, the CHY formalism leads to a very natural construction of the DDM form BCJ numerators: just fully expand the first $\Pfp(\Psi)$ and settle all the $\sigma_i$'s into Parke-Taylor factors:
\begin{equation}
    \Pfp(\Psi)\times\Pfp(\Psi)=\sum_{\pmb{\beta}\in S_{n-2}}n^{\text{YM}}\left(1,\pmb{\beta},n\right)\PT\left(1,\pmb{\beta},n\right)\Pfp(\Psi)\,.
\end{equation}
The main subject of this note is to provide a set of very straightforward graphic rules to read out these numerators directly, based on two previous papers~\cite{Fu:2017uzt,Teng:2017tbo}.\footnote{A recursive algorithm using another method is given in~\cite{Bjerrum-Bohr:2016axv}.} Interestingly, the single trace EYM integrands appear as intermediate steps in this construction. Suppose we have $m$ gravitons and $s=n-m$ gluons among these $n$ particles, then the EYM integrand is given by:
\begin{equation}
    \mathcal{I}_{\text{EYM}}(g_1g_2\ldots g_s\,|\,h_1\ldots h_m)=\PT\left(g_1g_2\ldots g_s\right)\Pf\left(\Psi_{\mathsf{H}}\right)\Pfp(\Psi)
\end{equation}
with $\mathsf{H}\equiv\{h_1h_2\ldots h_m\}$. After integrating over the CHY measure $d\Omega_{\text{CHY}}$, we obtain the corresponding amplitudes:
\begin{align}
\label{eq:CHYintegration}
    & A_{n}^{\text{GR}}(12\ldots n)=\int{d\Omega_{\text{CHY}}}\mathcal{I}_{\text{GR}}(12\ldots n)\nonumber\\
    & A_{s,m}^{\text{EYM}}(g_1\ldots g_s\,|\,h_1\ldots h_m)=(-1)^{\frac{(n+1)(n+2)}{2}+\frac{m(m+1)}{2}}\int{d\Omega_{\text{CHY}}}\mathcal{I}_{\text{EYM}}(g_1\ldots g_s\,|\,h_1\ldots h_m)\nonumber\\
    & A_{n}^{\text{YM}}(12\ldots n)=(-1)^{\frac{(n+1)(n+2)}{2}}\int{d\Omega_{\text{CHY}}}\mathcal{I}_{\text{YM}}(12\ldots n)\,.
\end{align}
In the above equations, the phase factors are meticulously chosen such that the resultant BCJ numerator has the simplest overall sign convention. The explicit construction of $n^{\text{YM}}$ will be presented in Section~\ref{sec:rules}, while the expansion of $\Pfp(\Psi)$ will be given in the next subsection.

The second set of double copy construction appears between $n$-point flavor ordered NLSM and the special Galileon integrand:
\begin{align}
    \mathcal{I}_{\text{NLSM}}=\PT(12\ldots)\left[\Pfp(A)\right]^{2}\,,& &\mathcal{I}_{\text{Galileon}}=\left[\Pfp(A)\right]^{4}\,.
\end{align}
Again, a well-controlled expansion of $\left[\Pfp(A)\right]^{2}$ can lead to
\begin{equation}
    \left[\Pfp(A)\right]^{4}=\sum_{\pmb\beta\in S_{\{2\ldots n-2\}}}n^{\text{NLSM}}\left(1,\pmb\beta,n\right)\PT(1,\pmb\beta,n)\left[\Pfp(A)\right]^{2}\,.
\end{equation}
In the intermediate steps of such expansion, one will encounter the $\text{NLSM}+\phi^{3}$ integrands~\cite{Cachazo:2016njl}:
\begin{equation}
\label{eq:NLSMphi3}
    \mathcal{I}_{\text{NLSM}+\phi^{3}}=\PT(12\ldots n)\,\PT(1,\pmb\alpha,n)\left[\Pf(A_{\overline{\pmb\alpha}})\right]^{2}\,,
\end{equation}
where $\pmb\alpha$ is an ordered subset of $\{2\ldots n-1\}$, and $\overline{\pmb\alpha}$ is the complement of $\pmb\alpha$ in $\{2\ldots n-1\}$. Physically, $\{1,\pmb\alpha,n\}$ is the set of bi-adjoint scalars and $\overline{\pmb\alpha}$ is the set of adjoint scalars. On the other hand, the same construction can be directly obtained from a dimensional reduction of the YM case. Both methods will be discussed in Section~\ref{sec:NLSM}

\subsection{Expansion of reduced Pfaffian}
If we choose $i=1$ and $j=n$ in Eq.~\eqref{eq:Pfp} for the deleted rows and columns, we can show that $\Pfp(\Psi)$ has the following expansion:
\begin{align}
\label{eq:expansion}
    \Pfp(\Psi)\cong(-1)^{\frac{(n+1)(n+2)}{2}}\sum_{\text{split}}\;\sum_{\pmb\alpha\in S_{\mathsf{A}}}W{(1,\pmb\alpha,n)}(-1)^{\frac{m(m+1)}{2}}(-1)^{m}\,\PT\left(1,\pmb{\alpha},n\right)\Pf(\Psi_{\mathsf{H}})\,.
\end{align}
By using $\cong$, we emphasize that this identity only holds when momentum conservation, transversality and scattering equation are all imposed. Now we describe the symbols used in Eq.~\eqref{eq:expansion}:
\begin{itemize}
    \item The position of particle $1$ and $n$ are fixed, which defines the KK basis.
    \item $\sum_{\text{split}}$ sums over all possible ways of splitting the set $\{2\ldots n-1\}$ into two subsets $\mathsf{A}$ and $\mathsf{H}$ (both of them can be empty). We use $m=|\mathsf{H}|$ to stand for the graviton number and $|\mathsf{A}|+2=s+2$ for the gluon number, such that $s+m=n-2$.
    \item $\sum_{\pmb{\alpha}\in S_{A}}$ sums over all possible gluon orderings.
    \item The symbol $W$ stands for a chain
    \begin{equation}
        W{(1,\pmb{\alpha},n)}=\epsilon_{1}\cdot F_{\alpha(1)}\cdot F_{\alpha(2)}\ldots F_{\alpha(s)}\cdot \epsilon_{n}\,,
    \end{equation}
    where $(F_i)^{\mu\nu}=(k_i)^{\mu}(\epsilon_{i})^{\nu}-(k_i)^{\nu}(\epsilon_{i})^{\mu}$ is the field strength tensor.
\end{itemize}
We can multiply both sides of Eq.~\eqref{eq:expansion} by another $\Pfp(\Psi)$ and then perform the CHY integration as in Eq.~\eqref{eq:CHYintegration}. What we get is the amplitude relation:
\begin{align}
\label{eq:GREYM}
    A_{n}^{\text{GR}}(12\ldots n)=\sum_{\text{split}}\;\sum_{\pmb\alpha\in S_{\mathsf{A}}}W(1,\pmb{\alpha},n)(-1)^{|\mathsf{H}|}A_{s+2,m}^{\text{EYM}}(1,\pmb{\alpha},n\,|\,\mathsf{H})\,.
\end{align}
This expansion was first derived in~\cite{Fu:2017uzt} by using Lam and Yao's cycle expansion~\cite{Lam:2016tlk}. In Appendix~\ref{sec:laplace}, we provide another proof using the Laplace expansion.

Eq.~\eqref{eq:GREYM} is an important intermediate step for deriving the BCJ numerator. The next job is thus to expand the EYM amplitudes in terms of pure YM ones. %In~\cite{Teng:2017tbo}, such an algorithm was derived for arbitrary number of gravitons.
In particular, we have
\begin{equation}
\label{eq:EYMexpansion}
    A_{s+2,m}^{\text{EYM}}(1,\pmb{\alpha},n\,|\,\mathsf{H})=\sum_{\pmb{\sigma}\in S_{\mathsf{H}}}\sum_{\pmb\gamma\in\pmb\alpha\shuffle\pmb\sigma} C_{\pmb\rho}\left(\pmb\sigma\right)A_{n}^{\text{YM}}(1,\pmb{\gamma},n)\,,
\end{equation}
where $\pmb{\alpha}\shuffle\pmb\sigma$ is the shuffle product of the set $\pmb\alpha$ and $\pmb\sigma$. The coefficients $C_{\pmb\rho}(\pmb\sigma)$, depending on a reference graviton order $\pmb\rho$, can then be evaluated from a set of graphic rules. Together with the origin of $\pmb\rho$, these rules are described in details in~\cite{Teng:2017tbo}. We note that the coefficients $C_{\pmb\rho}(\pmb\sigma)$ do not enjoy the explicit graviton permutation invariance. Of course, this symmetry can be restored by using certain combinations of BCJ relations.

Based on this result, we can easily give a set of graphic rules to directly construct the DDM form BCJ numerators for YM, which is the subject of the next section.

%%%%%%%%%%%%%%%%%%%%%%%%%%%%%%%%%%%%%%%%%%
\section{Graphic rules for BCJ numerators: YM}\label{sec:rules}
%%%%%%%%%%%%%%%%%%%%%%%%%%%%%%%%%%%%%%%%%%
As shown in \cite{Fu:2017uzt}, if we combine Eq.~\eqref{eq:GREYM} with \eqref{eq:EYMexpansion}, we get
\begin{align}
\label{eq:BCJconstruction}
    A_{n}^{\text{GR}}(12\ldots n)&=\sum_{\text{split}}\;\sum_{\pmb\alpha\in S_{\mathsf{A}}}\sum_{\pmb\sigma\in S_{\mathsf{H}}}\sum_{\pmb\gamma\in\pmb\alpha\shuffle\pmb\sigma}(-1)^{|\mathsf{H}|}W(1,\pmb\alpha,n)C_{\pmb\rho}(\pmb\sigma)A_{n}^{\text{YM}}(1,\pmb\gamma,n)\nonumber\\
    &=\sum_{\pmb\beta\in S_{\{2\ldots n-2\}}}n^{\text{YM}}(1,\pmb\beta,n)A_{n}^{\text{YM}}(1,\pmb\beta,n)\,.
\end{align}
Namely, by rearranging the sum in the first line, we can read out the desired DDM form BCJ numerator for tree level YM. Now given an order $\pmb\beta\in S_{n-2}$, we summarize the rearrangement into the following rules, based on spanning trees, for a direct evaluation of $n(1,\pmb\beta,n)$:

\paragraph{Step 1: Constructing the trees that contribute to the order $\pmb\beta$.} We first construct all the $n$-point \emph{increasing trees}\footnote{In an increasing tree, if we have two vertices $a$ and $b$ located on a path originating at the root, then $b$ cannot be closer to the root than $a$ if $a\prec b$.} with respect to the order $$1\prec\beta(1)\prec\beta(2)\prec\ldots\prec\beta(n-1)\prec n\,.$$ There are in all $(n-1)!$ such trees. All these trees are rooted on particle $1$, and contain $n$ as a leaf. %Next, we draw a path from $n$ to $1$ in each tree.

~\\
Eq.~\eqref{eq:BCJconstruction} implies that $\pmb\beta$ comes from a shuffle product $\pmb\alpha\shuffle\pmb\sigma$, and our next job is to identify ``gluons'' and ``gravitons'' in each tree, since they are treated differently according to Eq.~\eqref{eq:BCJconstruction}: gluons contribute a chain $W(1,\pmb\alpha,n)$ while gravitons give the coefficients $C_{\pmb\rho}(\pmb\sigma)$.

\paragraph{Step 2: Separation of gluons and gravitons.} We draw a path that originates at the root $1$ and ends at the leaf $n$. If this path contains $s$ particles besides $1$ and $n$, we denote it as $\mathcal{G}=\{1,\alpha(1),\alpha(2)\ldots\alpha(s),n\}$. Then we identify those in $\mathcal{G}$ as gluons and the rest $\mathsf{H}=\{h_1,h_2\ldots h_m\}$ as gravitons. Now the previously constructed spanning trees can be viewed as a set of planted graviton forests on the gluon roots $\{1,\alpha(1),\alpha(2)\ldots\alpha(s)\}$.

\paragraph{Step 3: Evaluation of graphs.} Now we present the algorithm to evaluate each graph constructed in the previous two steps:
\begin{itemize}
    \item For the gluon chain $\mathcal{G}=\{1,\alpha(1)\ldots\alpha(s),n\}$, we assign a factor:
    \begin{equation}
        W(1,\pmb\alpha,n)=\epsilon_1\cdot F_{\alpha(1)}\cdot F_{\alpha(2)}\cdot\ldots\cdot F_{\alpha(s)}\cdot\epsilon_n\,.
    \end{equation}
    Namely, we replace gluon $1$ and $n$ by the polarization $\epsilon_1$ and $\epsilon_n$, and for the rest gluons in $\mathcal{G}$, we replace them by the field strength tensor $F$.
    \item We choose a reference order $\pmb\rho$ for the graviton set $\mathsf{H}$,\footnote{We note that we have to use the same $\pmb\rho$ for a given set $\mathsf{H}$, even if this $\mathsf{H}$ appears in the evaluation of two different $n(1,\pmb\beta,n)$.} and then locate the position of $\rho(1)$ in the graph, from which we draw a path towards the gluon roots. Suppose this path ends at $\alpha(i)$, we can represent it by
    \begin{equation}
        \mathcal{P}\left[1\right]=\{\phi(1),\phi(2)\ldots\phi(\ell),V_1\}\,,
    \end{equation}
    with $\phi(1)=\rho(1)$ and $V_1=\alpha(i)$.
    \item Next, we delete $\mathcal{P}\left[1\right]$ from $\pmb\rho$ and construct the second path in the remaining vertices by the same way. We locate $\widetilde{\rho}(1)$ in
    \begin{equation*}
        \widetilde{\pmb\rho}\equiv\pmb\rho\backslash\{\phi(1),\phi(2)\ldots\phi(\ell)\}
    \end{equation*}
    and draw a path towards the gluon roots. Now this path can either ends on a gluon or on a previous traversed graviton. We call this endpoint $V_2$ for both cases. Then we can represent this path by
    \begin{equation}
        \mathcal{P}\left[2\right]=\{\widetilde{\phi}(1),\widetilde{\phi}(2)\ldots\widetilde{\phi}(t),V_2\}\,,
    \end{equation}
    with $\widetilde{\phi}(1)=\widetilde{\rho}(1)$. Repeat this process until we use up all the gravitons in $\pmb\rho$.
    \item Finally, we assign each graviton path a factor. For the first vertex in each path, we replace it by the corresponding polarization $\epsilon$; for the vertices in the middle of a path, we replace it by the corresponding field strength $F$; for the end point, we replace it by the corresponding momentum $k$, be it a gluon or graviton. For example, the first two paths $\mathcal{P}\left[1\right]$ and $\mathcal{P}\left[2\right]$ lead to
    \begin{align}
        \mathcal{P}\left[1\right]: & &&\epsilon_{\phi(1)}\cdot F_{\phi(2)}\cdot\ldots\cdot F_{\phi(\ell)}\cdot k_{V_1} \nonumber\\
        \mathcal{P}\left[2\right]: & &&\epsilon_{\widetilde{\phi}(1)}\cdot F_{\widetilde{\phi}(2)}\cdot\ldots\cdot F_{\widetilde{\phi}(t)}\cdot k_{V_2}\,,
    \end{align}
    where $V_1=\alpha(i)$ is a gluon, but $V_2$ can be either a gluon or a graviton in $\{\phi(1)\ldots\phi(\ell)\}$. As the last step, we multiply all these factors together with a phase $(-1)^{|\mathsf{H}|}$. Then $n^{\text{YM}}(1,\pmb\beta,n)$ is just the sum of these trees.
\end{itemize}
In the next two subsections, we explicitly evaluate some four-point and five-point BCJ numerators in the DDM basis, as examples.

\subsection{Four-point numerators}
For $n=4$, we only need to calculate $\pmb\beta=\{23\}$ and $\pmb\beta=\{32\}$. Then according to {\bf Step 1} and {\bf Step 2}, both of them are contributed by six trees. These trees are shown in Figure~\ref{fig:4p}, with gluon chains highlighted in red. Below each tree, we give the evaluation of each tree according to {\bf Step 3}. For $\mathsf{H}=\{23\}$, we use with the reference order $\pmb\rho=\{32\}$. Now summing over the trees, we get our results:
\begin{align}
\label{eq:n1234}
    n^{\text{YM}}(1,\{23\},4)&=(\epsilon_1\cdot\epsilon_4)\left[(\epsilon_3\cdot F_2\cdot k_1)+(\epsilon_3\cdot k_1)(\epsilon_2\cdot k_1)\right]+(\epsilon_1\cdot F_2\cdot F_3\cdot\epsilon_4)\nonumber\\
    &\quad-(\epsilon_1\cdot F_2\cdot\epsilon_4)\left[\epsilon_3\cdot (k_1+k_2)\right]-(\epsilon_1\cdot F_3\cdot\epsilon_4)(\epsilon_2\cdot k_1)\,,\\
\label{eq:n1324}
    n^{\text{YM}}(1,\{32\},4)&=(\epsilon_1\cdot\epsilon_4)\left[(\epsilon_3\cdot k_1)(\epsilon_2\cdot k_3)+(\epsilon_3\cdot k_1)(\epsilon_2\cdot k_1)\right]+(\epsilon_1\cdot F_3\cdot F_2\cdot\epsilon_4)\nonumber\\
    &\quad-(\epsilon_1\cdot F_3\cdot\epsilon_4)\left[\epsilon_2\cdot (k_1+k_3)\right]-(\epsilon_1\cdot F_2\cdot\epsilon_4)(\epsilon_3\cdot k_1)\,.
\end{align}
Up to a total sign caused by different conventions, the above numerators agree with those given in \cite{Fu:2017uzt}. The four-point gravity amplitude in this representation:
\begin{equation}
    A_{4}^{\text{GR}}(1234)=n^{\text{YM}}(1,\{23\},4)A_{4}^{\text{YM}}(1234)+n^{\text{YM}}(1,\{32\},4)A_{4}^{\text{YM}}(1324)
\end{equation}
does not have the manifest permutation invariance in particle $2$ and $3$. To be specific, the breaking appears in the first term of Eq.~\eqref{eq:n1234} and \eqref{eq:n1324}. However, the invariance can be easily restored by the BCJ relation:
\begin{equation}
    (k_1\cdot k_2)A_{4}^{\text{YM}}(1234)=(k_1\cdot k_3)A_{4}^{\text{YM}}(1324)\,.
\end{equation}

\begin{figure}[t]
    \centering
    \begin{tikzpicture}
    \filldraw [thick] (0,0) -- ++(-1,1) circle (2pt) node[above=1pt]{$2$};
    \filldraw [thick] (0,0) -- ++(0,1) circle (2pt) node[above=1pt]{$3$};
    \coordinate (d) at (1,1);
    \filldraw [red,thick] (0,0) circle (2pt) node[below=1pt]{$1$} -- ++(1,1) circle (2pt) node[above=1pt]{$4$};
    \node at (0,0) [below=0.5cm,font=\fontsize{7}{6}\selectfont]{$(\epsilon_1\cdot\epsilon_4)(\epsilon_3\cdot k_1)(\epsilon_2\cdot k_1)$};
    \filldraw [thick] (3.5,0) -- ++(-1,1) circle (2pt) node[above=1pt]{$2$} -- ++(1,0) circle (2pt) node[above=1pt]{$3$};
    \filldraw [thick,red] (3.5,0) circle (2pt) node[below=1pt]{$1$} -- ++(1,1) circle (2pt) node[above=1pt]{$4$};
    \node at (3.5,0) [below=0.5cm,font=\fontsize{7}{6}\selectfont]{$(\epsilon_1\cdot\epsilon_4)(\epsilon_3\cdot F_2\cdot k_1)$};
    \filldraw [thick] (0,-3) -- ++(-1,1) circle (2pt) node[above=1pt]{$2$};
    \filldraw [thick,red] (0,-3) circle (2pt) node[below=1pt]{$1$} -- ++(0,1) circle (2pt) node[above=1pt]{$3$} -- ++(1,0) circle (2pt) node[above=1pt]{$4$};
    \node at (0,-3) [below=0.5cm,font=\fontsize{7}{6}\selectfont]{$-(\epsilon_1\cdot F_3\cdot\epsilon_4)(\epsilon_2\cdot k_1)$};
    \filldraw [thick,red] (3.5,-3) circle (2pt) node[below=1pt]{$1$} -- ++(-1,1) circle (2pt) node[above=1pt]{$2$} -- ++(1,0) circle (2pt) node[above=1pt]{$3$} -- ++(1,0) circle (2pt) node[above=1pt]{$4$};
    \node at (3.5,-3) [below=0.5cm,font=\fontsize{7}{6}\selectfont]{$(\epsilon_1\cdot F_2\cdot F_3\cdot\epsilon_4)$};
    \filldraw [thick] (0,-6) -- ++(0,1) circle (2pt) node[above=1pt]{$3$};
    \filldraw [thick,red] (0,-6) circle (2pt) node[below=1pt]{$1$} -- ++(-1,1) circle (2pt) node[above=1pt]{$2$};
    \draw [thick,red] (0,-6) ++(-1,1) .. controls (0,-4.3) and (0,-4.3) .. ++(2,0);
    \fill [red] (0,-6) ++(1,1) circle (2pt) node[above=1pt]{$4$};
    \node at (0,-6) [below=0.5cm,font=\fontsize{7}{6}\selectfont]{$-(\epsilon_1\cdot F_2\cdot\epsilon_4)(\epsilon_3\cdot k_1)$};
    \filldraw [thick] (3.5,-6) ++(0,1) circle (2pt) node[above=1pt]{$3$} -- ++(-1,0);
    \filldraw [thick,red] (3.5,-6) circle (2pt) node[below=1pt]{$1$} -- ++(-1,1) circle (2pt) node[above=1pt]{$2$};
    \draw [thick,red] (3.5,-6) ++(-1,1) .. controls (3.5,-4.3) and (3.5,-4.3) .. ++(2,0);
    \fill [red] (3.5,-6) ++(1,1) circle (2pt) node[above=1pt]{$4$};
    \node at (3.5,-6) [below=0.5cm,font=\fontsize{7}{6}\selectfont]{$-(\epsilon_1\cdot F_2\cdot\epsilon_4)(\epsilon_3\cdot k_2)$};
    \draw [dashed,very thick] (5.25,2) -- ++(0,-9);
    \node at (1.75,2.) []{$n^{\text{YM}}(1,\{23\},4)$};
    %%%%%%%%%%%%%%%%%%%%%%%%%%%%%%%%%%%%%%%%%%%%%%%%%%%%%%%%%%%%%
    \begin{scope}[xshift=7cm]
    \node at (1.75,2.) []{$n^{\text{YM}}(1,\{32\},4)$};
    \filldraw [thick] (0,0) -- ++(-1,1) circle (2pt) node[above=1pt]{$2$};
    \filldraw [thick] (0,0) -- ++(0,1) circle (2pt) node[above=1pt]{$3$};
    \coordinate (d) at (1,1);
    \filldraw [red,thick] (0,0) circle (2pt) node[below=1pt]{$1$} -- ++(1,1) circle (2pt) node[above=1pt]{$4$};
    \node at (0,0) [below=0.5cm,font=\fontsize{7}{6}\selectfont]{$(\epsilon_1\cdot\epsilon_4)(\epsilon_3\cdot k_1)(\epsilon_2\cdot k_1)$};
    \filldraw [thick] (3.5,0) -- ++(0,1) circle (2pt) node[above=1pt]{$3$} -- ++(-1,0) circle (2pt) node[above=1pt]{$2$};
    \filldraw [thick,red] (3.5,0) circle (2pt) node[below=1pt]{$1$} -- ++(1,1) circle (2pt) node[above=1pt]{$4$};
    \node at (3.5,0) [below=0.5cm,font=\fontsize{7}{6}\selectfont]{$(\epsilon_1\cdot\epsilon_4)(\epsilon_3\cdot k_1)(\epsilon_2\cdot k_3)$};
    \filldraw [thick] (0,-3) -- ++(0,1) circle (2pt) node[above=1pt]{$3$};
    \filldraw [thick,red] (0,-3) circle (2pt) node[below=1pt]{$1$} -- ++(-1,1) circle (2pt) node[above=1pt]{$2$};
    \draw [thick,red] (0,-3) ++(-1,1) .. controls (0,-1.3) and (0,-1.3) .. ++(2,0);
    \fill [red] (1,-2) circle (2pt) node[above=1pt]{$4$};
    \node at (0,-3) [below=0.5cm,font=\fontsize{7}{6}\selectfont]{$-(\epsilon_1\cdot F_2\cdot\epsilon_4)(\epsilon_3\cdot k_1)$};
    \filldraw [thick,red] (3.5,-3) circle (2pt) node[below=1pt]{$1$} -- ++(0,1) circle (2pt) node[above=1pt]{$3$} -- ++(-1,0) circle (2pt) node[above=1pt]{$2$};
    \draw [thick,red] (3.5,-3) ++(-1,1) .. controls (3.5,-1.3) and (3.5,-1.3) .. ++(2,0);
    \fill [red] (3.5,-3) ++(1,1) circle (2pt) node[above=1pt]{$4$};
    \node at (3.5,-3) [below=0.5cm,font=\fontsize{7}{6}\selectfont]{$(\epsilon_1\cdot F_3\cdot F_2\cdot\epsilon_4)$};
    \filldraw [thick] (0,-6) -- ++(-1,1) circle (2pt) node[above=1pt]{$2$};
    \filldraw [thick,red] (0,-6) circle (2pt) node[below=1pt]{$1$} -- ++(0,1) circle (2pt) node[above=1pt]{$3$} -- ++(1,0) circle (2pt) node[above=1pt]{$4$};
    \node at (0,-6) [below=0.5cm,font=\fontsize{7}{6}\selectfont]{$-(\epsilon_1\cdot F_3\cdot\epsilon_4)(\epsilon_2\cdot k_1)$};
    \filldraw [thick] (3.5,-6) ++(-1,1) circle (2pt) node[above=1pt]{$2$} -- ++(1,0);
    \filldraw [thick,red] (3.5,-6) circle (2pt) node[below=1pt]{$1$} -- ++(0,1) circle (2pt) node[above=1pt]{$3$} -- ++(1,0) circle (2pt) node[above=1pt]{$4$};
    \node at (3.5,-6) [below=0.5cm,font=\fontsize{7}{6}\selectfont]{$-(\epsilon_1\cdot F_3\cdot\epsilon_4)(\epsilon_2\cdot k_3)$};
    \end{scope}
    \end{tikzpicture}
    \caption{The spanning trees corresponding to $n^{\text{YM}}(1,\{23\},4)$ and $n^{\text{YM}}(1,\{32\},4)$. The evaluation is performed under the reference order $\pmb\rho=\{32\}$.}
    \label{fig:4p}
\end{figure}
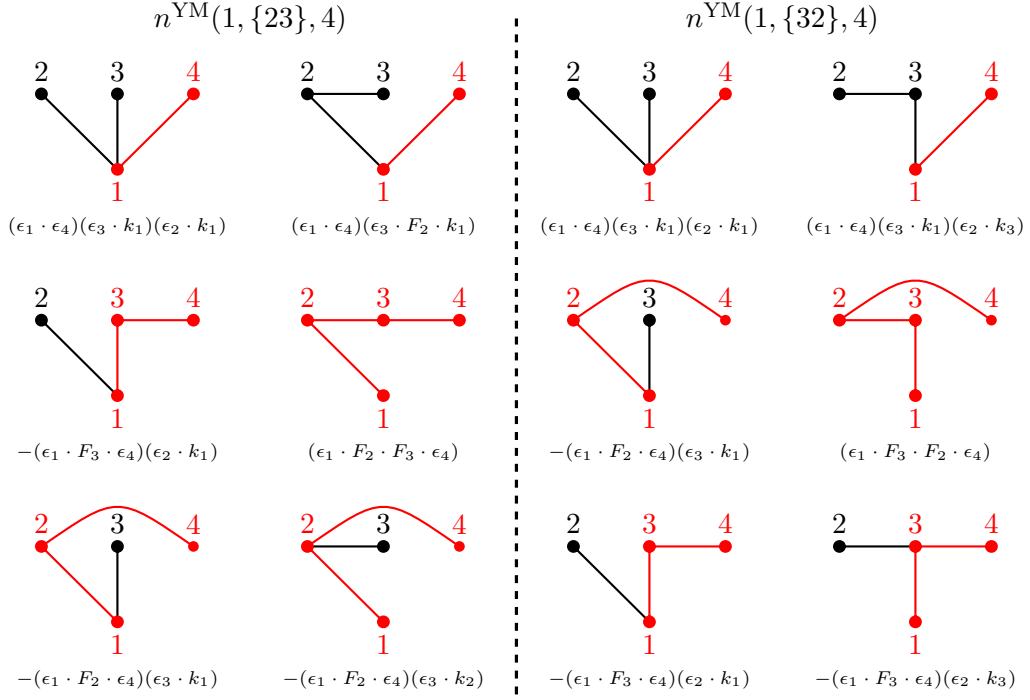

\begin{figure}[t]
    \centering
    \begin{tikzpicture}[starnode/.style={text=blue,font=\Huge}]
        \begin{scope}
            \coordinate (a) at (0,0);
            \filldraw [thick] (a) -- ++(-1.5,1) circle (2pt) node[above=1pt]{$2$} (a) -- ++(-0.5,1) circle(2pt) node[above=1pt]{$3$} (a) -- ++(0.5,1) circle (2pt) node[above=1pt]{$4$};
            \filldraw [thick,red] (a) circle (2pt) node[below=1pt]{$1$} -- ++(1.5,1) circle (2pt) node[above=1pt]{$5$};
        \end{scope}
        %%%%%%%%%%%%%%%%%%%%%%%%%%%%%%%%%%%%%%%%%%%%%%
        \begin{scope}[xshift=3.8cm]
            \coordinate (a) at (0,0);
            \filldraw [thick] (a) -- ++(-1.5,1) circle (2pt) node[above=1pt]{$2$} (a) -- ++(-0.5,1) circle(2pt) node[above=1pt]{$3$};
            \filldraw [thick,red] (a) circle  (2pt) node[below=1pt]{$1$} -- ++(0.5,1) circle (2pt) node[above=1pt]{$4$} -- ++(1,0) circle (2pt) node[above=1pt]{$5$};
        \end{scope}
        %%%%%%%%%%%%%%%%%%%%%%%%%%%%%%%%%%%%%%%%%%%%%%
        \begin{scope}[xshift=7.6cm]
            \coordinate (a) at (0,0);
            \filldraw [thick] (a) -- ++(-1.5,1) circle (2pt) node[above=1pt]{$2$} (a) -- ++(0.5,1) circle (2pt) node[above=1pt]{$4$};
            \filldraw [red,thick] (a) circle (2pt) node[below=1pt]{$1$} -- ++(-0.5,1) circle (2pt) node [above=1pt]{$3$} (a) ++(1.5,1) circle (2pt) node [above=1pt]{$5$};
            \draw [thick,red] (-0.5,1) .. controls (0.5,1.7) and (0.5,1.7) .. (1.5,1);
        \end{scope}
        %%%%%%%%%%%%%%%%%%%%%%%%%%%%%%%%%%%%%%%%%%%%%%
        \begin{scope}[xshift=11.4cm]
            \coordinate (a) at (0,0);
            \filldraw [thick] (a) -- ++(-0.5,1) circle (2pt) node[above=1pt]{$3$} (a) -- ++(0.5,1) circle (2pt) node[above=1pt]{$4$};
            \filldraw [red,thick] (a) circle (2pt) node[below=1pt]{$1$} -- ++(-1.5,1) circle (2pt) node[above=1pt]{$2$} (a) ++(1.5,1) circle (2pt) node[above=1pt]{$5$};
            \draw [thick,red] (-1.5,1) .. controls (0,1.8) and (0,1.8) .. (1.5,1);
        \end{scope}
        %%%%%%%%%%%%%%%%%%%%%%%%%%%%%%%%%%%%%%%%%%%%%%
        \begin{scope}[xshift=0cm,yshift=-2.2cm]
            \coordinate (a) at (0,0);
            \filldraw [thick] (a) -- ++(-1.5,1) circle (2pt) node[above=1pt]{$2$} -- ++(1,0) circle (2pt) node[above=1pt]{$3$} (a) -- ++(0.5,1) circle (2pt) node[above=1pt]{$4$};
            \filldraw [red,thick] (a) circle (2pt) node[below=1pt]{$1$} -- ++(1.5,1) circle (2pt) node[above=1pt]{$5$};
        \end{scope}
        %%%%%%%%%%%%%%%%%%%%%%%%%%%%%%%%%%%%%%%%%%%%%%
        \begin{scope}[xshift=3.8cm,yshift=-2.2cm]
            \coordinate (a) at (0,0);
            \filldraw [thick] (a) -- ++(-1.5,1) circle (2pt) node[above=1pt]{$2$} -- ++(1,0) circle (2pt) node[above=1pt]{$3$};
            \filldraw [red,thick] (a) circle (2pt) node[below=1pt]{$1$} -- ++(0.5,1) circle (2pt) node[above=1pt]{$4$} -- ++(1,0) circle (2pt) node[above=1pt]{$5$};
            %\node at (0.75,0) [starnode]{$\star$};
        \end{scope}
        %%%%%%%%%%%%%%%%%%%%%%%%%%%%%%%%%%%%%%%%%%%%%%
        \begin{scope}[xshift=7.6cm,yshift=-2.2cm]
            \coordinate (a) at (0,0);
            \filldraw [thick] (a) -- ++(0.5,1) circle (2pt) node[above=1pt]{$4$};
            \filldraw [red,thick] (a) circle (2pt) node[below=1pt]{$1$} -- ++(-1.5,1) circle (2pt) node[above=1pt]{$2$} -- ++(1,0) circle (2pt) node[above=1pt]{$3$} (a) ++(1.5,1) circle (2pt) node[above=1pt]{$5$};
            \draw [red,thick] (-0.5,1) .. controls (0.5,1.7) and (0.5,1.7) .. (1.5,1);
        \end{scope}
        %%%%%%%%%%%%%%%%%%%%%%%%%%%%%%%%%%%%%%%%%%%%%%
        \begin{scope}[xshift=11.4cm,yshift=-2.2cm]
            \coordinate (a) at (0,0);
            \filldraw [thick] (a) -- ++(0.5,1) circle (2pt) node[above=1pt]{$4$} ++ (-1,0) circle (2pt) node[above=1pt]{$3$} -- ++(-1,0);
            \filldraw [red,thick] (a) circle (2pt) node[below=1pt]{$1$} -- ++(-1.5,1) circle (2pt) node[above=1pt]{$2$} (a) ++(1.5,1) circle (2pt) node[above=1pt]{$5$};
            \draw [thick,red] (-1.5,1) .. controls (0,1.8) and (0,1.8) .. (1.5,1);
        \end{scope}
        %%%%%%%%%%%%%%%%%%%%%%%%%%%%%%%%%%%%%%%%%%%%%
        \begin{scope}[xshift=0cm,yshift=-4.4cm]
            \coordinate (a) at (0,0);
            \filldraw [thick] (a) -- ++(-0.5,1) circle (2pt) node[above=1pt]{$3$} -- ++(1,0) circle (2pt) node[above=1pt]{$4$} (a) -- ++(-1.5,1) circle (2pt) node[above=1pt]{$2$};
            \filldraw [thick,red] (a) circle (2pt) node[below=1pt]{$1$} -- ++(1.5,1) circle (2pt) node[above=1pt]{$5$};
        \end{scope}
        %%%%%%%%%%%%%%%%%%%%%%%%%%%%%%%%%%%%%%%%%%%%%
        \begin{scope}[xshift=3.8cm,yshift=-4.4cm]
            \coordinate (a) at (0,0);
            \filldraw [thick] (a) -- ++(-1.5,1) circle (2pt) node[above=1pt]{$2$};
            \filldraw [thick,red] (a) circle (2pt) node[below=1pt]{$1$} -- ++(-0.5,1) circle (2pt) node[above=1pt]{$3$} -- ++(1,0) circle (2pt) node[above=1pt]{$4$} -- ++(1,0) circle (2pt) node[above=1pt]{$5$};
        \end{scope}
        %%%%%%%%%%%%%%%%%%%%%%%%%%%%%%%%%%%%%%%%%%%%%
        \begin{scope}[xshift=7.6cm,yshift=-4.4cm]
            \coordinate (a) at (0,0);
            \filldraw [thick] (a) -- ++(-1.5,1) circle (2pt) node[above=1pt]{$2$} ++(1,0) -- ++(1,0) circle (2pt) node[above=1pt]{$4$};
            \filldraw [thick,red] (a) circle (2pt) node[below=1pt]{$1$} -- ++(-0.5,1) circle (2pt) node[above=1pt]{$3$} ++(2,0) circle (2pt) node[above=1pt]{$5$};
            \draw [thick,red] (-0.5,1) .. controls (0.5,1.7) and (0.5,1.7) .. (1.5,1);
        \end{scope}
        %%%%%%%%%%%%%%%%%%%%%%%%%%%%%%%%%%%%%%%%%%%%%
        \begin{scope}[xshift=11.4cm,yshift=-4.4cm]
            \coordinate (a) at (0,0);
            \filldraw [thick] (a) -- ++(-0.5,1) circle (2pt) node[above=1pt]{$3$} -- ++(1,0) circle (2pt) node[above=1pt]{$4$};
            \filldraw [red,thick] (a) circle (2pt) node[below=1pt]{$1$} -- ++(-1.5,1) circle (2pt) node[above=1pt]{$2$} ++ (3,0) circle (2pt) node[above=1pt]{$5$};
            \draw [thick,red] (-1.5,1) .. controls (0,1.8) and (0,1.8) .. (1.5,1);
            %\node at (0.75,0) [starnode]{$\star$};
        \end{scope}
        %%%%%%%%%%%%%%%%%%%%%%%%%%%%%%%%%%%%%%%%%%%%%
        \begin{scope}[xshift=0cm,yshift=-6.6cm]
            \coordinate (a) at (0,0);
            \filldraw [thick] (a) -- ++(-1.5,1) circle (2pt) node[above=1pt]{$2$} -- ++(1,0) circle (2pt) node[above=1pt]{$3$} -- ++(1,0) circle (2pt) node[above=1pt]{$4$};
            \filldraw [thick,red] (a) circle (2pt) node[below=1pt]{$1$} -- ++(1.5,1) circle (2pt) node[above=1pt]{$5$};
        \end{scope}
        %%%%%%%%%%%%%%%%%%%%%%%%%%%%%%%%%%%%%%%%%%%%%
        \begin{scope}[xshift=3.8cm,yshift=-6.6cm]
            \coordinate (a) at (0,0);
            \filldraw [thick,red] (a) circle (2pt) node[below=1pt]{$1$} -- ++(-1.5,1) circle (2pt) node[above=1pt]{$2$} -- ++(1,0) circle (2pt) node[above=1pt]{$3$} -- ++(1,0) circle (2pt) node[above=1pt]{$4$} -- ++(1,0) circle (2pt) node[above=1pt]{$5$};
            %\node at (0.75,0) [starnode]{$\star$};
        \end{scope}
        %%%%%%%%%%%%%%%%%%%%%%%%%%%%%%%%%%%%%%%%%%%%%
        \begin{scope}[xshift=7.6cm,yshift=-6.6cm]
            \coordinate (a) at (0,0);
            \filldraw [thick] (0.5,1) circle (2pt) node[above=1pt]{$4$} -- ++(-1,0);
            \filldraw [red,thick] (a) circle (2pt) node[below=1pt]{$1$} -- ++(-1.5,1) circle (2pt) node[above=1pt]{$2$} -- ++(1,0) circle (2pt) node[above=1pt]{$3$} (a) ++(1.5,1) circle (2pt) node[above=1pt]{$5$};
            \draw [red,thick] (-0.5,1) .. controls (0.5,1.7) and (0.5,1.7) .. (1.5,1);
        \end{scope}
        %%%%%%%%%%%%%%%%%%%%%%%%%%%%%%%%%%%%%%%%%%%%%
        \begin{scope}[xshift=11.4cm,yshift=-6.6cm]
            \coordinate (a) at (0,0);
            \filldraw [thick] (0.5,1) circle (2pt) node[above=1pt]{$4$} -- ++(-1,0) circle (2pt) node[above=1pt]{$3$} -- ++(-1,0);
            \filldraw [thick,red] (a) circle (2pt) node[below=1pt]{$1$} -- ++(-1.5,1) circle (2pt) node[above=1pt]{$2$} ++(3,0) circle (2pt) node[above=1pt]{$5$};
            \draw [thick,red] (-1.5,1) .. controls (0,1.8) and (0,1.8) .. (1.5,1);
            %\node at (0.75,0) [starnode]{$\star$};
        \end{scope}
        %%%%%%%%%%%%%%%%%%%%%%%%%%%%%%%%%%%%%%%%%%%%%
        \begin{scope}[xshift=0cm,yshift=-8.8cm]
            \coordinate (a) at (0,0);
            \filldraw [thick] (a) -- ++(-1.5,1) circle (2pt) node[above=1pt]{$2$} (a) -- ++(-0.5,1) circle (2pt) node[above=1pt]{$3$} (a) ++(0.5,1) circle (2pt) node[above=1pt]{$4$};
            \draw [thick] (-1.5,1) .. controls (-0.5,1.8) and (-0.5,1.8) .. (0.5,1);
            \filldraw [red,thick] (a) circle (2pt) node[below=1pt]{$1$} -- ++(1.5,1) circle (2pt) node[above=1pt]{$5$};
        \end{scope}
        %%%%%%%%%%%%%%%%%%%%%%%%%%%%%%%%%%%%%%%%%%%%%
        \begin{scope}[xshift=3.8cm,yshift=-8.8cm]
            \coordinate (a) at (0,0);
            \filldraw [thick] (a) -- ++(-0.5,1) circle (2pt) node[above=1pt]{$3$};
            \filldraw [red,thick] (a) circle (2pt) node[below=1pt]{$1$} -- ++(-1.5,1) circle (2pt) node[above=1pt]{$2$} (a) ++(1.5,1) circle (2pt) node[above=1pt]{$5$} -- ++(-1,0) circle (2pt) node[above=1pt]{$4$};
            \draw [thick,red] (-1.5,1) .. controls (-0.5,1.8) and (-0.5,1.8) .. (0.5,1);
        \end{scope}
        %%%%%%%%%%%%%%%%%%%%%%%%%%%%%%%%%%%%%%%%%%%%%
        \begin{scope}[xshift=7.6cm,yshift=-8.8cm]
            \coordinate (a) at (0,0);
            \filldraw [thick] (a) -- ++(-1.5,1) circle (2pt) node[above=1pt]{$2$} (a) ++(0.5,1) circle (2pt) node[above=1pt]{$4$};
            \draw [thick] (-1.5,1) .. controls (-0.5,1.8) and (-0.5,1.8) .. (0.5,1);
            \filldraw [thick,red] (a) circle (2pt) node[below=1pt]{$1$} -- ++(-0.5,1) circle (2pt) node[above=1pt]{$3$} (a) ++(1.5,1) circle (2pt) node[above=1pt]{$5$};
            \draw [thick,red] (-0.5,1) .. controls (0.5,1.8) and (0.5,1.8) .. (1.5,1);
            %\node at (0.75,0) [starnode]{$\star$};
        \end{scope}
        %%%%%%%%%%%%%%%%%%%%%%%%%%%%%%%%%%%%%%%%%%%%
        \begin{scope}[xshift=11.4cm,yshift=-8.8cm]
            \coordinate (a) at (0,0);
            \filldraw [thick] (a) -- ++(-0.5,1) circle (2pt) node[above=1pt]{$3$} (a) ++(0.5,1) circle (2pt) node[above=1pt]{$4$};
            \draw [thick] (-1.5,1) .. controls (-0.5,1.8) and (-0.5,1.8) .. (0.5,1);
            \filldraw [red,thick] (a) circle (2pt) node[below=1pt]{$1$} -- ++(-1.5,1) circle (2pt) node[above=1pt]{$2$} (a) ++(1.5,1) circle (2pt) node[above=1pt]{$5$};
            \draw [thick,red] (-1.5,1) .. controls (0,1.8) and (0,1.8) .. (1.5,1);
        \end{scope}
        %%%%%%%%%%%%%%%%%%%%%%%%%%%%%%%%%%%%%%%%%%
        \begin{scope}[xshift=0cm,yshift=-11cm]
            \coordinate (a) at (0,0);
            \filldraw [thick] (a) -- ++(-1.5,1) circle (2pt) node[above=1pt]{$2$} -- ++(1,0) circle (2pt) node[above=1pt]{$3$} (a) ++(0.5,1) circle (2pt) node[above=1pt]{$4$};
            \draw [thick] (-1.5,1) .. controls (-0.5,1.8) and (-0.5,1.8) .. (0.5,1);
            \filldraw [red,thick] (a) circle (2pt) node[below=1pt]{$1$} -- ++(1.5,1) circle (2pt) node[above=1pt]{$5$};
        \end{scope}
        %%%%%%%%%%%%%%%%%%%%%%%%%%%%%%%%%%%%%%%%%%%%%
        \begin{scope}[xshift=3.8cm,yshift=-11cm]
            \coordinate (a) at (0,0);
            \filldraw [thick] (-1.5,1) -- ++(1,0) circle (2pt) node[above=1pt]{$3$};
            \filldraw [red,thick] (a) circle (2pt) node[below=1pt]{$1$} -- ++(-1.5,1) circle (2pt) node[above=1pt]{$2$} (a) ++(1.5,1) circle (2pt) node[above=1pt]{$5$} -- ++(-1,0) circle (2pt) node[above=1pt]{$4$};
            \draw [thick,red] (-1.5,1) .. controls (-0.5,1.8) and (-0.5,1.8) .. (0.5,1);
            %\node at (0.75,0) [text=gray,font=\Huge]{$\star$};
        \end{scope}
        %%%%%%%%%%%%%%%%%%%%%%%%%%%%%%%%%%%%%%%%%%%%%
        \begin{scope}[xshift=7.6cm,yshift=-11cm]
            \coordinate (a) at (0,0);
            \filldraw [thick] (a) ++(0.5,1) circle (2pt) node[above=1pt]{$4$};
            \draw [thick] (-1.5,1) .. controls (-0.5,1.8) and (-0.5,1.8) .. (0.5,1);
            \filldraw [thick,red] (a) circle (2pt) node[below=1pt]{$1$} -- ++(-1.5,1) circle (2pt) node[above=1pt]{$2$} circle (2pt) -- ++(1,0) circle (2pt) node[above=1pt]{$3$} (a) ++(1.5,1) circle (2pt) node[above=1pt]{$5$};
            \draw [thick,red] (-0.5,1) .. controls (0.5,1.8) and (0.5,1.8) .. (1.5,1);
            %\node at (0.75,0) [text=gray,font=\Huge]{$\star$};
        \end{scope}
        %%%%%%%%%%%%%%%%%%%%%%%%%%%%%%%%%%%%%%%%%%%%
        \begin{scope}[xshift=11.4cm,yshift=-11cm]
            \coordinate (a) at (0,0);
            \filldraw [thick] (-0.5,1) circle (2pt) node[above=1pt]{$3$} -- ++(-1,0) (a) ++(0.5,1) circle (2pt) node[above=1pt]{$4$};
            \draw [thick] (-1.5,1) .. controls (-0.5,1.8) and (-0.5,1.8) .. (0.5,1);
            \filldraw [red,thick] (a) circle (2pt) node[below=1pt]{$1$} -- ++(-1.5,1) circle (2pt) node[above=1pt]{$2$} (a) ++(1.5,1) circle (2pt) node[above=1pt]{$5$};
            \draw [thick,red] (-1.5,1) .. controls (0,1.8) and (0,1.8) .. (1.5,1);
        \end{scope}
    \end{tikzpicture}
    \caption{The spanning trees corresponding to $n(1,\{234\},5)$ for YM. All gluon vertices are highlighted in red. %Those trees with a star have the right shapes that possibly contribute to $6$-point NLSM numerators. Among them, the blue starred ones contribute to $n(1,\{2345\},6)$.
    }
    \label{fig:5p}
\end{figure}
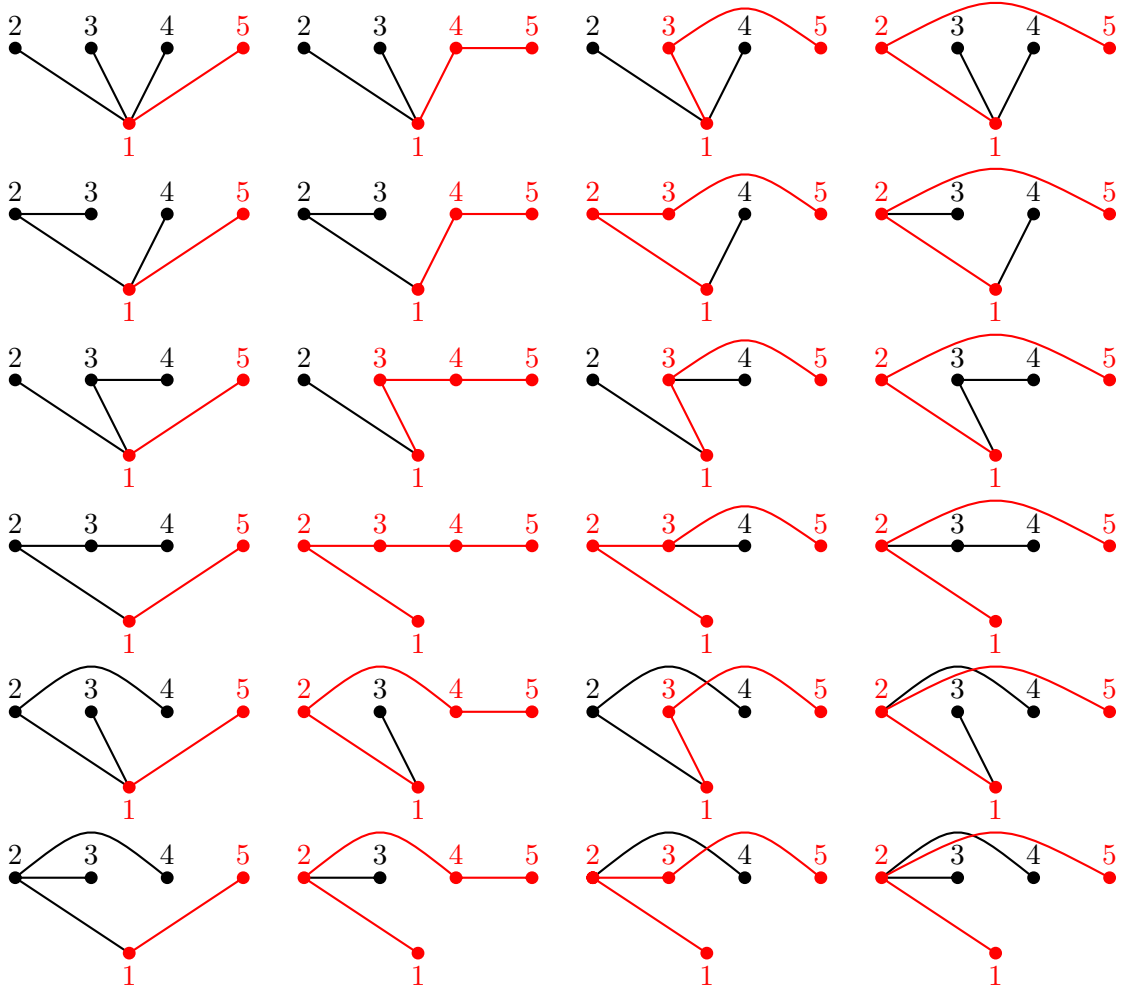

\subsection{Five-point numerators}\label{sec:5p}
For $n=5$, we need to evaluate six numerators, corresponding to $\pmb\beta\in S_{\{234\}}$. Among them, we only give the explicit calculation of $\pmb\beta=\{234\}$. We list all the spanning trees that contribute to this order in Figure~\ref{fig:5p}, with gluon chains highlighted in red. After evaluating all these graphs in Figure~\ref{fig:5p}, we obtain the result:\footnote{For each $\mathsf{H}$, we always choose the reference order $\pmb\rho$ as the descending order of the elements in $\mathsf{H}$.}
\begingroup
\allowdisplaybreaks
\begin{align}
    \tt{row1}&=-(\epsilon_1\cdot\epsilon_5)(\epsilon_4\cdot k_1)(\epsilon_3\cdot k_1)(\epsilon_2\cdot k_1)+(\epsilon_1\cdot F_4\cdot\epsilon_5)(\epsilon_3\cdot k_1)(\epsilon_2\cdot k_1)\nonumber\\*
    &\quad+(\epsilon_1\cdot F_3\cdot \epsilon_5)(\epsilon_4\cdot k_1)(\epsilon_2\cdot k_1)+(\epsilon_1\cdot F_2\cdot \epsilon_5)(\epsilon_4\cdot k_1)(\epsilon_3\cdot k_1)\\
    \tt{row2}&=-(\epsilon_1\cdot\epsilon_5)(\epsilon_4\cdot k_1)(\epsilon_3\cdot F_2\cdot k_1)+(\epsilon_1\cdot F_4\cdot\epsilon_5)(\epsilon_3\cdot F_2\cdot k_1)\nonumber\\*
    &\quad-(\epsilon_1\cdot F_2\cdot F_3\cdot \epsilon_5)(\epsilon_4\cdot k_1)+(\epsilon_1\cdot F_2\cdot\epsilon_5)(\epsilon_4\cdot k_1)(\epsilon_3\cdot k_2)\\
    \tt{row3}&=-(\epsilon_1\cdot\epsilon_5)(\epsilon_4\cdot F_3\cdot k_1)(\epsilon_2\cdot k_1)-(\epsilon_1\cdot F_3\cdot F_4\cdot\epsilon_5)(\epsilon_2\cdot k_1)\nonumber\\*
    &\quad+(\epsilon_1\cdot F_3\cdot\epsilon_5)(\epsilon_4\cdot k_3)(\epsilon_2\cdot k_1)+(\epsilon_1\cdot F_2\cdot\epsilon_5)(\epsilon_4\cdot F_3\cdot k_1)\\
    \tt{row4}&=-(\epsilon_1\cdot\epsilon_5)(\epsilon_4\cdot F_3\cdot F_2\cdot k_1)+(\epsilon_1\cdot F_2\cdot F_3\cdot F_4\cdot\epsilon_5)\nonumber\\*
    &\quad-(\epsilon_1\cdot F_2\cdot F_3\cdot\epsilon_5)(\epsilon_4\cdot k_3)+(\epsilon_1\cdot F_2\cdot\epsilon_5)(\epsilon_4\cdot F_3\cdot k_2)\\
    \tt{row5}&=-(\epsilon_1\cdot\epsilon_5)(\epsilon_4\cdot F_2\cdot k_1)(\epsilon_3\cdot k_1)-(\epsilon_1\cdot F_2\cdot F_4\cdot\epsilon_5)(\epsilon_3\cdot k_1)\nonumber\\*
    &\quad+(\epsilon_1\cdot F_3\cdot\epsilon_5)(\epsilon_4\cdot F_2\cdot k_1)+(\epsilon_1\cdot F_2\cdot\epsilon_5)(\epsilon_4\cdot k_2)(\epsilon_3\cdot k_1)\\
    \tt{row6}&=-(\epsilon_1\cdot\epsilon_5)(\epsilon_4\cdot F_2\cdot k_1)(\epsilon_3\cdot k_2)-(\epsilon_1\cdot F_2\cdot F_4\cdot\epsilon_5)(\epsilon_3\cdot k_2)\nonumber\\*
    &\quad-(\epsilon_1\cdot F_2\cdot F_3\cdot\epsilon_5)(\epsilon_4\cdot k_2)+(\epsilon_1\cdot F_2\cdot\epsilon_5)(\epsilon_4\cdot k_2)(\epsilon_3\cdot k_2)\,.
\end{align}
\endgroup
The numerator $n(1,\{234\},5)$ is thus the sum of these graphs:
\begin{equation}
    n^{\text{YM}}(1,\{234\},5)=\tt{row1}+\tt{row2}+\tt{row3}+\tt{row4}+\tt{row5}+\tt{row6}\,.
\end{equation}
All the other numerators can be evaluated similarly, and we put their expressions in Appendix~\ref{sec:numerator}.

%%%%%%%%%%%%%%%%%%%%%%%%%%%%%%%%%%%%%%%%%%%%%%%%%%%%
\section{Graphic rules for BCJ numerators: NLSM}\label{sec:NLSM}
%%%%%%%%%%%%%%%%%%%%%%%%%%%%%%%%%%%%%%%%%%%%%%%%%%%%

In this section, we give a set of graphic rules for a direct evaluation of DDM form BCJ numerators for NLSM. It has been proposed first in~\cite{Cachazo:2014xea} that the CHY integrand for NLSM can be obtained through a dimensional reduction from a higher dimensional YM. In this work, we follow another dimensional reduction scheme given in~\cite{Huang:2017ydz}, which enable us to derive immediately the graphic rules to construct DDM form BCJ numerator for NLSM from those for YM.

%%%%%%%%%%%%%%%%%%%%%%%%%%%%%%%%%%%%%%%%%%%%%%%%%%%%%%%%%%
\subsection{Dimensional reduction}\label{sec:reduction}
%%%%%%%%%%%%%%%%%%%%%%%%%%%%%%%%%%%%%%%%%%%%%%%%%%%%%%%%%%

We start with a YM integrand in $(d+d+d)$-dimensions and construct the matrix $\Psi$ out of the following momenta and polarizations:
\begin{align}
    K_{a}=(k_a;0;0)& &\mathcal{E}_{a}=\left\{\begin{array}{>{\displaystyle}l @{\hspace{1.5em}} >{\displaystyle}l}
    (0;0;\epsilon_a) & a=1\text{ and }n \\
    (0;\epsilon_a;0) & a=2\ldots n-1
    \end{array}\right.\,,
\end{align}
where both $k_a$ and $\epsilon_a$ are in $d$-dimensions. We can then construct the matrices $\mathcal{A}$, $\mathcal{B}$, $\mathcal{C}$ according to Eq.~\eqref{eq:ABC}, and the matrix $\Psi^{(d+d+d)}$ as:
\begin{equation}
    \Psi^{(d+d+d)}=\left(\begin{array}{cc}
    \mathcal{A} & -\mathcal{C}^{T} \\
    \mathcal{C} & \mathcal{B}
    \end{array}\right)\,.
\end{equation}
Obviously, we have $\mathcal{C}=0$ and $\mathcal{A}=A$.\footnote{We reserve $A$, $B$ and $C$ to denote those matrices appearing in $d$-dimensional integrands.} Consequently, the reduced Pfaffian of $\Psi^{(d+d+d)}$ factorizes into:
\begin{equation}
    \Pfp\big[\Psi^{(d+d+d)}\big]=\Pfp(A)\Pf(\mathcal{B})=\frac{(-1)^{n}\epsilon_{1}\cdot\epsilon_n}{\sigma_{1n}}\,\Pfp(A)\,\Pf(B_{1,n}^{1,n})\,.
\end{equation}
Next, we make the replacement $\epsilon_a\rightarrow k_a$ in the above equation, which leads to:
\begin{equation}
\label{eq:DR}
    \left.\Pfp\big[\Psi^{(d+d+d)}\big]\right|_{\epsilon_a\rightarrow k_a}=-(k_1\cdot k_n)\left[\Pfp(A)\right]^{2}\,.
\end{equation}
This dimensional reduction implies that if we perform the following replacement
\begin{align}
\label{eq:NLSMreplacement}
    &\epsilon_a\cdot k_b\;\rightarrow\;0 \nonumber\\ &\epsilon_a\cdot\epsilon_b\;\rightarrow\;\left\{\begin{array}{>{\displaystyle}l @{\hspace{1.5em}} >{\displaystyle}l}
    k_a\cdot k_b & \{a,b\}\subset\{2\ldots n-1\}\text{ or }\{a,b\}=\{1,n\}\\
    0 & a\in\{1,n\}\text{ and }b\in\{2\ldots n-1\}\,\text{, or vice versa}
    \end{array}\right.
\end{align}
directly in Eq.~\eqref{eq:expansion}, the expansion of $d$-dimensional YM integrand $\Pfp(\Psi)$, we get the expansion of the $d$-dimensional NLSM integrand $[\Pfp(A)]^2$ in the KK basis.

%%%%%%%%%%%%%%%%%%%%%%%%%%%%%%%%%%%%%%%%%%%%%%%%%%%%%%%%%%
\subsection{Recursive expansion}\label{sec:NLSMrecursion}
%%%%%%%%%%%%%%%%%%%%%%%%%%%%%%%%%%%%%%%%%%%%%%%%%%%%%%%%%%

As an alternative approach, we can derive the expansion of $[\Pfp(A)]^2$ from a recursive relation. Such a recursion can be proved by the Laplace expansion of $\Pfp(A)$. The object central to this calculation is
\begin{equation}
    \PT(1,\pmb\alpha,n)\left[\Pf(A_{\overline{\pmb\alpha}})\right]^{2}\,.
\end{equation}
This object is just the $\mathcal{I}_{R}$ for $\text{NLSM}+\phi^{3}$, the soft extension of NLSM~\cite{Cachazo:2016njl}. The definition of the symbols involved can be found in Eq.~\eqref{eq:NLSMphi3}. For the simplest case, there are only two adjoint scalars, namely, $\overline{\pmb\alpha}=\{qp\}$. Then we have
\begin{align}
\label{eq:2adj}
    \PT(1,\pmb\alpha,n)\left[\Pf(A_{\{qp\}})\right]^{2}&=\PT(1,\pmb\alpha,n)\left(\frac{k_q\cdot k_p}{\sigma_{qp}}\right)^{2}\nonumber\\
    &\cong-\PT(1,\pmb\alpha,n)\left(\frac{k_1\cdot k_q}{\sigma_{1q}}+\sum_{i}\frac{k_{\alpha_i}\cdot k_q}{\sigma_{\alpha_iq}}+\frac{k_n\cdot k_q}{\sigma_{nq}}\right)\frac{k_q\cdot k_p}{\sigma_{pq}}\nonumber\\
    &\cong-\left(\frac{k_p\cdot k_q}{\sigma_{pq}}\right)\sum_{\shuffle}\left(k_q\cdot Y_{q}\right)\PT(1,\pmb\alpha\shuffle\{q\},n)+\frac{(k_p\cdot k_q)^{2}}{\sigma_{nq}\sigma_{pq}}\,\PT(1,\pmb\alpha,n)\,,
\end{align}
where $Y_q$, which implicitly depends on the orderings contained in the shuffle product $\pmb\alpha\shuffle\{q\}$, is the sum of the momenta of bi-adjoint scalars ahead of the adjoint scalar $q$ in each ordering. To obtain the last equity, we have used the momentum conservation and the identity:
\begin{equation}
    \frac{\PT(1,\pmb\alpha,n)}{\sigma_{\alpha_iq}}=\PT(1\ldots\alpha_i,q,\alpha_{i+1}\ldots n)+\frac{\PT(1,\pmb\alpha,n)}{\sigma_{\alpha_{i+1}q}}\,.
\end{equation}
Now by moving the last term in Eq.~\eqref{eq:2adj} to the left hand side of the equation, we can solve that:
\begin{align}
    \PT(1,\pmb\alpha,n)\left[\Pf(A_{\{qp\}})\right]^{2}&\cong\PT(1,\pmb\alpha,n)\left(\frac{k_q\cdot k_p}{\sigma_{qp}}\right)^{2}\cong\left(k_p\cdot k_q\right)\sum_{\shuffle}\left(k_q\cdot Y_{q}\right)\PT(1,\pmb\alpha\shuffle\{q\},n)\,.
\end{align}
To deal with generic cases, we first define the following recursive relation:
\begin{align}
    &\quad U^{\mu}\left[1,\pmb\alpha\shuffle\{s_m s_{m+1}\ldots s_{i}\},n\,|\,\{s_{m-1}\ldots s_1\}\right]\nonumber\\
    &=\sum_{\shuffle}(Y_{s_m})^{\mu}\left[\Pf(A_{\{s_{m-1}\ldots s_{1}\}})\right]^{2}\PT(1,\pmb\alpha\shuffle\{s_{m}\ldots s_{i}\},n)\nonumber\\
    &\quad+\sum_{\substack{\ell,t=1 \\ \ell\neq t}}^{m-1}\left(k_{s_\ell}\right)^{\mu}\left(k_{s_\ell}\cdot k_{s_t}\right)k_{s_t}\cdot U\left[1,\pmb\alpha\shuffle\{s_ts_ls_m\ldots s_i\},n\,|\,\{s_{m-1}\ldots\slashed{s}_{\ell}\ldots\slashed{s}_{t}\ldots s_1\}\right]\,.
\end{align}
This recursion finally lands on $U\left[\ldots|\varnothing\right]$ or $U\left[\ldots|\{s_1\}\right]$, which are defined as:
\begin{align}
    & U^{\mu}\left[1,\pmb\alpha\shuffle\{s_1\ldots s_i\},n\,|\,\varnothing\right]=\sum_{\shuffle}(Y_{s_1})^{\mu}\PT(1,\pmb\alpha\shuffle\{s_1\ldots s_i\},n)\nonumber\\
    & U^{\mu}\left[1,\pmb\alpha\shuffle\{s_2\ldots s_i\},n\,|\,\{s_1\}\right]=0
\end{align}
The second condition guarantees that when the right list of $U^{\mu}$ has an odd number of elements, we have $U^{\mu}=0$ identically. Finally, we can express the recursive expansion of generic $\text{NLSM}+\phi^{3}$ integrand:
\begin{align}
    \PT(1,\pmb\alpha,n)\left[\Pf(A_{\{s_1s_2\ldots s_m\}})\right]^{2}\cong\sum_{t=1}^{m-1}\left(k_{s_m}\cdot k_{s_t}\right)k_{s_t}\cdot U\left[1,\pmb\alpha\shuffle\{s_ts_m\},n\,|\,\{s_{m-1}\ldots\slashed{s}_t\ldots s_1\}\right]\,.
\end{align}
The proof is straightforward using the various identites shown in~\cite{Teng:2017tbo}. Finally, the pure NLSM corresponds to $\pmb\alpha=\varnothing$ in the above equation, such that we have
\begin{align}
    \left[\Pfp(A)\right]^{2}&\cong-\PT(1n)\left[\Pf(A_{1,n}^{1,n})\right]^{2}\nonumber\\
    &\cong-\sum_{t=2}^{n-2}\left(k_{n-1}\cdot k_{t}\right)k_{t}\cdot U\left[1,\varnothing\shuffle\{t,n-1\},n\,|\,\{n-2\ldots\slashed{t}\ldots 2\}\right]\,,
\end{align}
which enables us to expand the NLSM integrand in terms of the $\text{NLSM}+\phi^{3}$ and pure bi-adjoint $\phi^{3}$ integrands.

%%%%%%%%%%%%%%%%%%%%%%%%%%%%%%%%%%%%%%%%%%%%%%%%%%%%%%%%%%
\subsection{Graphic rules}\label{sec:NLSMrules}
%%%%%%%%%%%%%%%%%%%%%%%%%%%%%%%%%%%%%%%%%%%%%%%%%%%%%%%%%%

The above dimensional reduction scheme shown in Section~\ref{sec:reduction} implies that once we make the replacement \eqref{eq:NLSMreplacement} to the rules given in Section~\ref{sec:rules}, we immediately get the set of rules for the direct evaluation of the DDM form BCJ numerators for NLSM:
\begin{equation}
    \left[\Pfp(A)\right]^{2}=\sum_{\pmb\beta\in S_{\{2\ldots n-1\}}}n^{\text{NLSM}}\left(1,\pmb\beta,n\right)\PT(1,\pmb\beta,n)\,.
\end{equation}
We note that the same set of rules can also be derived from the recursion given in Section~\ref{sec:NLSMrecursion}.

We can still follow {\bf Step 1} and {\bf Step 2} in Section~\ref{sec:rules}, reaching the separation of ``gluons'' and ``gravtions''. Then in the evaluation of the ``gluon chain'' we have
\begin{equation}
    \epsilon_1\cdot F_{\alpha(1)}\cdot F_{\alpha(2)}\cdot\ldots\cdot F_{\alpha(s)}\cdot\epsilon_n\;\xrightarrow{\eqref{eq:NLSMreplacement}}\;\left\{\begin{array}{>{\displaystyle}c @{\hspace{1.5em}} >{\displaystyle}l}
    k_1\cdot k_n & \pmb\alpha=\varnothing \\
    0 & \text{otherwise}
    \end{array}\right.\,.
\end{equation}
Therefore, for NLSM, when constructing spanning trees, we only need to consider those with $n$ directly attached to the root $1$. On the other hand, the factor $k_1\cdot k_n$ is exactly the coefficient in front of $[\Pfp(A)]^2$ according to~\eqref{eq:DR}, so that for an evaluation of $[\Pfp(A)]^2$, we only need to evaluate the spanning trees of $\{1\ldots n-1\}$ rooted on $1$. Next, for a ``graviton chain'', we have:
\begin{equation*}
    \epsilon_{\phi(1)}\cdot F_{\phi(2)}\cdot\ldots\cdot F_{\phi(\ell)}\cdot k_{V_1}
\end{equation*}
for YM. Then under the replacement \eqref{eq:NLSMreplacement}, only the term:
\begin{equation*}
    (\epsilon_{\phi(1)}\cdot\epsilon_{\phi(2)})(k_{\phi_2}\cdot k_{\phi(3)})\ldots (\epsilon_{\phi(\ell-1)}\cdot\epsilon_{\phi(\ell)})(k_{\phi(\ell)}\cdot k_{V_1})
\end{equation*}
remains nonzero. Such terms only appear when $\ell$ is even, namely, the path contains an even number of edges. Therefore, we have:
\begin{equation}
\label{eq:NLSMchain}
    \epsilon_{\phi(1)}\cdot F_{\phi(2)}\cdot\ldots\cdot F_{\phi(\ell)}\cdot k_{V_1}\;\xrightarrow{\eqref{eq:NLSMreplacement}}\;
    (-1)^{\frac{\ell}{2}}(k_{\phi(1)}\cdot k_{\phi(2)})(k_{\phi(2)}\cdot k_{\phi(3)})\ldots(k_{\phi(\ell)}\cdot k_{V_1})
\end{equation}
when $\ell$ is even, and zero when $\ell$ is odd. Consequently, we only need to take into account those spanning trees whose graviton chains all contain an even number of edges.

Now we summarize the graphic rules for constructing the DDM form BCJ numerator $n^{\text{NLSM}}(1,\pmb\beta,n)$ for NLSM as the following:

\paragraph{Step 1: Constructing the trees that contribute to the order $\pmb\beta$.} We first construct all the $(n-1)$-point increasing trees with respect to the order:
\begin{equation*}
    1\prec\beta(1)\prec\beta(2)\prec\ldots\prec\beta(n-2)\,.
\end{equation*}
There are in all $(n-2)!$ such trees, all of which have $1$ as the root.

\paragraph{Step 2: Evaluation of graphs} We choose a reference order $\pmb\rho$ for the particles $\{2\ldots n-1\}$, and then locate the position of $\rho(1)$ in the graph, from which we draw a path towards the root $1$. We represent it as
\begin{equation}
    \mathcal{P}\left[1\right]=\{\phi(1),\phi(2)\ldots\phi(\ell),V_1\}\,,
\end{equation}
with $\phi(1)=\rho(1)$ and $V_1=1$. Now we have the following situations:
\begin{itemize}
    \item If this path contains an odd number of edges (i.e., $\ell$ is odd), then this entire tree does not contribute.
    \item If this path contains an even number of edges, we. delete $\mathcal{P}\left[1\right]$ from $\pmb\rho$. Then we construct another path in the same way from the remaining points. Namely, we locate $\widetilde{\rho}(1)$ in
    \begin{equation*}
        \widetilde{\pmb\rho}=\pmb\rho\backslash\{\phi(1)\ldots\phi(\ell)\}\,,
    \end{equation*}
    and draw a path towards the root $1$. Now this path, represented by $$\mathcal{P}\left[2\right]=\{\widetilde{\phi}(1),\widetilde{\phi}(2)\ldots\widetilde{\phi}(t),V_2\}\,,$$ may end either at the root: $V_2=1$, or another previously traversed point $V_2=\phi(i)$.
    \item Whenever we meet such a path $\mathcal{P}\left[i\right]$ with an odd number of edges, the entire tree does not contribute and we simply delete it.
\end{itemize}
Now we can assign factors to the remaining trees, in which all $\mathcal{P}\left[i\right]$ contains an even number of edges. For a given tree, we replace the end points of each path by $k$, and the internal vertices by $k^{\mu}k^{\nu}$. For example, we have:
\begin{align}
    \mathcal{P}\left[1\right]: & && (k_{\phi(1)}\cdot k_{\phi(2)})(k_{\phi(2)}\cdot k_{\phi(3)})\ldots (k_{\phi(\ell-1)}\cdot k_{\phi(\ell)})(k_{\phi(\ell)}\cdot k_{V_1})\nonumber\\
    \mathcal{P}\left[2\right]: & && (k_{\widetilde{\phi}(1)}\cdot k_{\widetilde{\phi}(2)})(k_{\widetilde{\phi}(2)}\cdot k_{\widetilde{\phi}(3)})\ldots (k_{\widetilde{\phi}(t-1)}\cdot k_{\widetilde{\phi}(t)})(k_{\widetilde{\phi}(t)}\cdot k_{V_2})\,,
\end{align}
where $V_1=1$, and $V_2$ can be either the root $1$, or one of the $\phi(i)$'s.\footnote{Notice that in our rules, we do not include the factor $(-1)^{\ell/2}$ as in Eq.~\eqref{eq:NLSMchain}. The reason is that for each of the surviving trees, such factors all multiply to an overall phase $(-1)^{\frac{n-2}{2}}$, where $n-2$ is exactly the number of edges in an $(n-1)$-point tree. For the same reason, we neglect the phase $(-1)^{|\mathsf{H}|}$ appearing in the YM rules, since it becomes another overall phase $(-1)^{n-2}$ in the NLSM case.} Finally, $n^{\text{NLSM}}(1,\pmb\beta,n)$ is obtained by summing up all these contributing trees.

\subsection{Examples}\label{sec:NLSMexample}
According our rules, we immediately see that when $n$ is odd, all these numerators vanish and so does the amplitude. The reason is that according to our rules, those trees that contribute to the numerators must have all $\mathcal{P}\left[i\right]$ with even edges. On the other hand, a spanning tree with $n-1$ vertex must have exactly $n-2$ edges. Therefore:
\begin{equation}
    \sum_{i}\left|\mathcal{P}[i]\right|=n-2=\text{even}\,,
\end{equation}
such that $n$ must be even in order to possibly have nonzero numerators.

\paragraph{Four point NLSM.} For $n=4$, we need to calculate $n(1,\{23\},4)$ and $n(1,\{32\},4)$. The spanning trees that possibly contribute are those in the first row of Figure~\ref{fig:4p}, with the red edge $(14)$ deleted. Among these four graphs, only one of them is nonzero under the reference order $\pmb\rho=\{32\}$:
\begin{equation}
    \adjustbox{raise=-0.75cm}{\begin{tikzpicture}
    \filldraw [thick] (0,0) circle (2pt) node[below=1pt]{$1$} -- ++(0,1) circle (2pt) node[above=1pt]{$2$} -- ++(1,0) circle (2pt) node[above=1pt]{$3$};
    \end{tikzpicture}}=(k_3\cdot k_2)(k_2\cdot k_1)\,.
\end{equation}
Therefore, when $n=4$, we have:
\begin{align}
\label{eq:NLSM4p}
    n^{\text{NLSM}}(1,\{23\},4)=(k_3\cdot k_2)(k_2\cdot k_1)\,,& && n^{\text{NLSM}}(1,\{32\},n)=0\,.
\end{align}

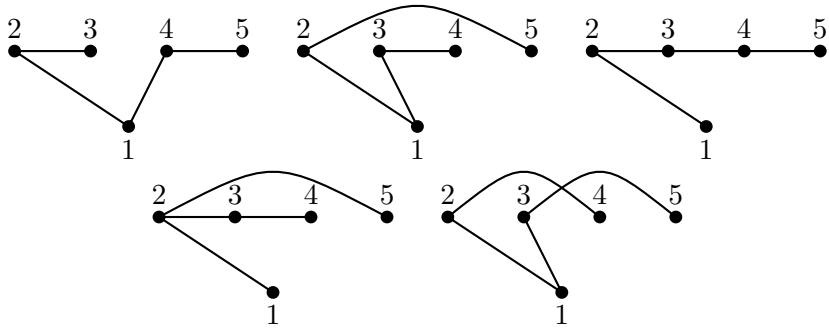
\begin{figure}[t]
\centering
    \begin{tikzpicture}[starnode/.style={text=blue,font=\Huge}]
    \begin{scope}
        \coordinate (a) at (0,0);
        \filldraw [thick] (a) -- ++(-1.5,1) circle (2pt) node[above=1pt]{$2$} -- ++(1,0) circle (2pt) node[above=1pt]{$3$};
        \filldraw [thick] (a) circle (2pt) node[below=1pt]{$1$} -- ++(0.5,1) circle (2pt) node[above=1pt]{$4$} -- ++(1,0) circle (2pt) node[above=1pt]{$5$};
    %    \node at (0.75,0) [starnode]{$\star$};
    \end{scope}
    \begin{scope}[xshift=3.8cm,yshift=0cm]
        \coordinate (a) at (0,0);
        \filldraw [thick] (a) -- ++(-0.5,1) circle (2pt) node[above=1pt]{$3$} -- ++(1,0) circle (2pt) node[above=1pt]{$4$};
        \filldraw [thick] (a) circle (2pt) node[below=1pt]{$1$} -- ++(-1.5,1) circle (2pt) node[above=1pt]{$2$} ++ (3,0) circle (2pt) node[above=1pt]{$5$};
        \draw [thick] (-1.5,1) .. controls (0,1.8) and (0,1.8) .. (1.5,1);
    %    \node at (0.75,0) [starnode]{$\star$};
    \end{scope}
    \begin{scope}[xshift=7.6cm,yshift=0cm]
        \coordinate (a) at (0,0);
        \filldraw [thick] (a) circle (2pt) node[below=1pt]{$1$} -- ++(-1.5,1) circle (2pt) node[above=1pt]{$2$} -- ++(1,0) circle (2pt) node[above=1pt]{$3$} -- ++(1,0) circle (2pt) node[above=1pt]{$4$} -- ++(1,0) circle (2pt) node[above=1pt]{$5$};
   %     \node at (0.75,0) [starnode]{$\star$};
    \end{scope}
    \begin{scope}[xshift=1.9cm,yshift=-2.2cm]
        \coordinate (a) at (0,0);
        \filldraw [thick] (0.5,1) circle (2pt) node[above=1pt]{$4$} -- ++(-1,0) circle (2pt) node[above=1pt]{$3$} -- ++(-1,0);
        \filldraw [thick] (a) circle (2pt) node[below=1pt]{$1$} -- ++(-1.5,1) circle (2pt) node[above=1pt]{$2$} ++(3,0) circle (2pt) node[above=1pt]{$5$};
        \draw [thick] (-1.5,1) .. controls (0,1.8) and (0,1.8) .. (1.5,1);
 %       \node at (0.75,0) [starnode]{$\star$};
    \end{scope}
    \begin{scope}[xshift=5.7cm,yshift=-2.2cm]
        \coordinate (a) at (0,0);
        \filldraw [thick] (a) -- ++(-1.5,1) circle (2pt) node[above=1pt]{$2$} (a) ++(0.5,1) circle (2pt) node[above=1pt]{$4$};
        \draw [thick] (-1.5,1) .. controls (-0.5,1.8) and (-0.5,1.8) .. (0.5,1);
        \filldraw [thick] (a) circle (2pt) node[below=1pt]{$1$} -- ++(-0.5,1) circle (2pt) node[above=1pt]{$3$} (a) ++(1.5,1) circle (2pt) node[above=1pt]{$5$};
        \draw [thick] (-0.5,1) .. controls (0.5,1.8) and (0.5,1.8) .. (1.5,1);
%        \node at (0.75,0) [starnode]{$\star$};
    \end{scope}
 %   \begin{scope}[xshift=5.7cm,yshift=-2.2cm]
 %       \coordinate (a) at (0,0);
 %       \filldraw [thick] (-1.5,1) -- ++(1,0) circle (2pt) node[above=1pt]{$3$};
 %       \filldraw [thick] (a) circle (2pt) node[below=1pt]{$1$} -- ++(-1.5,1) circle (2pt) node[above=1pt]{$2$} (a) ++(1.5,1) circle (2pt) node[above=1pt]{$5$} -- ++(-1,0) circle (2pt) node[above=1pt]{$4$};
 %       \draw [thick] (-1.5,1) .. controls (-0.5,1.8) and (-0.5,1.8) .. (0.5,1);
            %\node at (0.75,0) [text=gray,font=\Huge]{$\star$};
 %   \end{scope}
 %   \begin{scope}[xshift=9.5cm,yshift=-2.2cm]
 %       \coordinate (a) at (0,0);
 %       \filldraw [thick] (a) ++(0.5,1) circle (2pt) node[above=1pt]{$4$};
 %       \draw [thick] (-1.5,1) .. controls (-0.5,1.8) and (-0.5,1.8) .. (0.5,1);
 %       \filldraw [thick] (a) circle (2pt) node[below=1pt]{$1$} -- ++(-1.5,1) circle (2pt) node[above=1pt]{$2$} circle (2pt) -- ++(1,0) circle (2pt) node[above=1pt]{$3$} (a) ++(1.5,1) circle (2pt) node[above=1pt]{$5$};
 %       \draw [thick] (-0.5,1) .. controls (0.5,1.8) and (0.5,1.8) .. (1.5,1);
            %\node at (0.75,0) [text=gray,font=\Huge]{$\star$};
 %   \end{scope}
    \end{tikzpicture}
    \caption{The $5$-point spanning trees that contribute to $n^{\text{NLSM}}(1,\{2345\},6)$.}
    \label{fig:6p}
\end{figure}

\paragraph{Six point NLSM.} For $n=6$, we need to calculate $n(1,\pmb\beta,6)$ for $\pmb\beta\in S_{\{2345\}}$. There are in all $24$ of them. As an example, the spanning trees that contribute to $\pmb\beta=\{2345\}$ under the reference order $\pmb\rho=\{5432\}$ are shown in Figure~\ref{fig:6p}. Now following the rules given above, we have:
\begin{align}
    n^{\text{NLSM}}(1,\{2345\},6)&=(k_5\cdot k_4)(k_4\cdot k_1)(k_3\cdot k_2)(k_2\cdot k_1)+(k_5\cdot k_2)(k_2\cdot k_1)(k_4\cdot k_3)(k_3\cdot k_1)\nonumber\\
    &\quad+(k_5\cdot k_4)(k_4\cdot k_3)(k_3\cdot k_2)(k_2\cdot k_1)+(k_5\cdot k_2)(k_2\cdot k_1)(k_4\cdot k_3)(k_3\cdot k_2)\nonumber\\
    &\quad+(k_5\cdot k_3)(k_3\cdot k_1)(k_4\cdot k_2)(k_2\cdot k_1)\,.
\end{align}
All the other numerators can be evaluated in this way, and the results are shown in Appendix~\ref{sec:NLSMnum}.

\paragraph{Number of nonzero numerators.} It is interesting to note that following our rules, the number of nonzero numerators is \emph{at most} $(n-2)!-(n-3)!$. One may naively think that there are in all $(n-2)!$ nonzero numerators corresponding to all $(n-2)!$ permutations $\pmb\beta$ of the set $\{2\ldots n-1\}$. Actually, given a reference order $\pmb\rho=\{\rho(1),\rho(2)\ldots\rho(n-2)\}$, all the numerators with $\rho(n-2)$ coming first vanishes:
\begin{equation}
    n^{\text{NLSM}}(1,\rho(n-2),\widetilde{\pmb\beta} ,n)=0\,,\qquad\widetilde{\pmb\beta}\in S_{\{2\ldots n-2\}\backslash\rho(n-2)}\,,
\end{equation}
The reason is that all the increasing trees with respect to $\pmb\beta=\{\rho(n-2),\widetilde{\pmb\beta}\}$ must have $\rho(n-2)$ connected directly to the root $1$. Then according to our rules, there must exist an odd path $\mathcal{P}[1]=\{\rho(n-2),1\}$ contained in all these trees, such that no increasing tree of $\pmb\beta=\{\rho(n-2),\widetilde{\pmb\beta}\}$ contributes. This is reflected in our $6$-point example given in Appendix~\ref{sec:NLSMnum}, namely, all $n^{\text{NLSM}}(1,5,\pmb\beta(234),6)=0$. On the other hand, there are still other vanishing numerators. For $n=6$, we have in all $13$ nonzero numerators, comparing to $4!=24$ under naive expectation. Figuring out the exact number calls for a detailed study on the combinatorics of spanning trees, which is deferred to a future work.

%%%%%%%%%%%%%%%%%%%%%%%%%%%%%%%%%%%%%%%%%%%%%%%%%%%%%%%%%%%%%%%%
\subsection{The equivalence to other constructions}
%%%%%%%%%%%%%%%%%%%%%%%%%%%%%%%%%%%%%%%%%%%%%%%%%%%%%%%%%%%%%%%%

In \cite{Du:2016tbc} and \cite{Carrasco:2016ldy}, two different constructions are proposed from the off-shell extension of BCJ relation in NLSM and Abelian Z-theory respectively:
\begin{align}
\label{eq:DF}
    &\text{Du and Fu~\cite{Du:2016tbc}:} && n^{\text{DF}}(1,\pmb\beta,n)=\sum_{\pmb\rho}S[\pmb\beta|\pmb\rho], \\
\label{eq:CMS}
    &\text{Carrasco, Mafra and Schlotterer~\cite{Carrasco:2016ldy}:} && n^{\text{CMS}}(1,\pmb\beta,n)=(-1)^{\frac{n}{2}}S[\pmb\beta|\pmb\beta]\,,
\end{align}
 %
 %\bea
 %n^{(1)}_{1,\sigma({2,\cdots n-1}),n}&\equiv&\Sl_{\rho}S[\sigma|\rho]
 %\eea
 %and
 %\bea
 %n^{(2)}_{1,\sigma(2,\cdots,n-1),n}\equiv S[\sigma|\sigma],
 %\eea
 %
where $S$ is the momentum kernel for the KLT relation~\cite{Kawai:1985xq,Bern:1998ug,BjerrumBohr:2010ta,BjerrumBohr:2010zb,BjerrumBohr:2010yc,BjerrumBohr:2010hn}. The convention for $S$ follows the paper \cite{Carrasco:2016ldy}. In the DF construction, $\rho$ presents a certain subset of permutations (see \cite{Du:2016tbc}). Noticing that only amplitudes with even number external particles in NLSM are nonzero, so that the ${n/2}$ in the CMS construction must be an integer.  Both constructions manifest $(n-2)!$ permutation symmetry. In contrary, the numerators presented in this work has the following features:
\begin{itemize}
    \item They are polynomials of Mandelstam variables.
    \item The total number of nonzero numerators are less than $(n-2)!-(n-3)!$.
    \item The permutation invariance is not manifest.
\end{itemize}
Therefore, our construction in Section~\ref{sec:NLSMrules} essentially descends from \eqref{eq:DF} and \eqref{eq:CMS} by using a few BCJ relations in such a way that no poles are introduced. However, the price is that the manifest permutation invariance is lost.

Next, we consider the $4$-point case as an example for how to relate these $(n-2)!$ symmetric constructions to the numerators in this paper. The explicit numerators from \eqref{eq:DF} and \eqref{eq:CMS} are
\begin{align}
\label{eq:nDF}
    & n^{\text{DF}}(1,2,3,4)=S[23|32]=s_{21}s_{31}\,, && n^{\text{DF}}(1,3,2,4)=S[32|23]=s_{31}s_{21}\,, \\
\label{eq:nCMS}
    & n^{\text{CMS}}(1,2,3,4)=S[23|23]=s_{21}(s_{32}+s_{31})\,, && n^{\text{CMS}}(1,3,2,4)=S[32|32]=s_{31}(s_{23}+s_{21})\,.
\end{align}
 %
 %\bea
 %n^{(1)}_{1,2,3,4}=S[23|32]=s_{21}s_{31}, n^{(1)}_{1,3,2,4}=s_{31}s_{21}
 %\eea
 %and
 %\bea
 %n^{(2)}_{1,2,3,4}=S[23|23]=s_{21}(s_{32}+s_{31}), %n^{(2)}_{1,3,2,4}=S[32|32]=s_{31}(s_{23}+s_{21}).
 %\eea
 %
Using Eq.~\eqref{eq:nDF}, we write down the full amplitude as
\bea
\label{eq:Adf}
\mathcal{A}^{\text{NLSM}}(1,2,3,4)&=&s_{21}s_{31}A^{\phi^3}(1,2,3,4)+s_{31}s_{21} A^{\phi^3}(1,3,2,4)\nn
&=&s_{21}s_{31}A^{\phi^3}(1,2,3,4)-(s_{31}+s_{32})s_{21}A^{\phi^3}(1,2,3,4)\nn
&=&-s_{32}s_{21}A^{\phi^3}(1,2,3,4)\,,
\eea
where $A^{\phi^3}$ denotes the bi-adjoint $\phi^3$ amplitudes, and we have used the BCJ relation to obtain the last line. Up to a sign, which may caused by convention, and a few factors of $2$s coming from $s_{ab}=2k_a\cdot k_b$, the above expression is identical to the $4$-point construction~\eqref{eq:NLSM4p} in the this paper. The full amplitude from the second construction \eqref{eq:nCMS} gives
\bea
\label{eq:Acms}
\mathcal{A}^{\text{NLSM}}(1,2,3,4)&=&s_{21}(s_{32}+s_{31}) A^{\phi^3}(1,2,3,4)+s_{31}(s_{23}+s_{21}) A^{\phi^3}(1,3,2,4)\nn
&=&s_{21}(s_{32}+s_{31}) A^{\phi^3}(1,2,3,4)-s_{31}s_{21} A^{\phi^3}(1,3,2,4)\nn
&=&s_{32}s_{21} A^{\phi^3}(1,2,3,4)\,.
\eea
Again, we arrive at the construction in this paper.\footnote{Due to the double copy relation, we can replace simultaneously $\mathcal{A}^{\text{NLSM}}$ and $A^{\phi^3}$ in \eqref{eq:Adf} and \eqref{eq:Acms} by the special Galileon amplitude $A^{\text{Galileon}}$ and the flavor ordered NLSM amplitude $A^{\text{NLSM}}$.}

In order to understand more on the CHY formalism for NLSM, it is worthy to work out the full connection to the existing constructions derived from different ways. We leave this discussion to a future work.

%%%%%%%%%%%%%%%%%%%%%%%%%%%%%%%%%%%%%%%%%%%%%%%%%%%%%%%%%%
\section{Conclusion and discussion}\label{sec:conclusion}
%%%%%%%%%%%%%%%%%%%%%%%%%%%%%%%%%%%%%%%%%%%%%%%%%%%%%%%%%%

In this work, we have shown that the DDM form BCJ numerators for both YM and NLSM can be evaluated directly by a set of graphic rules, derived from the expansion of the reduced Pfaffians in the CHY integrands. These numerators are explicit polynomial functions of the scalar products involving $k$'s and $\epsilon$'s for YM, Mandelstam variables for NLSM.
\begin{itemize}
\item For YM, there are in all $(n-2)!$ such numerators $n^{\text{YM}}(1,\pmb\beta,n)$, each of which contains a sum of $(n-1)!$ terms, corresponding to the increasing trees with respect to the order $1\prec\beta(1)\prec\ldots\prec\beta(n-2)\prec n$. The evaluation of these trees depends on a reference order $\pmb\rho$, which makes the permutation invariance not manifest but significantly simplifies our results.

\item For NLSM, there are \emph{at most} $(n-2)!-(n-3)!$ such numerators $n^{\text{NLSM}}(1,\pmb\beta,n)$. Given a reference order $\pmb\rho=\{\rho(1),\rho(2)\ldots\rho(n-2)\}$, we have shown that all the $(n-3)!$ numerators $n^{\text{NLSM}}(1,\rho(n-2),\widetilde{\pmb\beta}, n)=0$. Each surviving numerator is contributed by the increasing trees of the order $1\prec\beta(1)\prec\ldots\prec\beta(n-2)$ that satisfy the following criteria: all the paths $\mathcal{P}[i]$ constructed according to the rules given in Section~\ref{sec:NLSMrules} have even length.
\end{itemize}
Finally, we have given a few explicit examples to illustrate our construction and discuss the connection with other existing constructions.

There are a number of future directions as suggested by our work. First of all, one can further investigate how the numerators constructed by our rules may give insight to manifest the kinematic algebra at the dynamical level, namely, in a Lagrangian. This goal has been achieved in NLSM by a field redefinition~\cite{Cheung:2016prv}, while similar approach has only been applied to the self-dual sector~\cite{Monteiro:2011pc} for YM.

Second, it is interesting to explore how to restore the manifest permutation invariance for YM, EYM and NLSM. One very straightforward approach is to average over the numerators obtained by all the $(n-2)!$ reference orders. The symmetrized version is more complicated and unwieldy. On the other hand, one can also prove the permutation invariance by repeatedly using of BCJ relations. Remarkably, the involved BCJ relations may not be in their usual forms: polarization vectors can get entangled with momenta~\cite{Chiodaroli:2017ngp}. Physically, it may indicate an interplay with the constraints imposed by gauge invariance~\cite{Arkani-Hamed:2016rak}.

Third, one can ask what is the underlying string theory that leads to the numerators we have constructed. For example, the form \eqref{eq:CMS} can be derived from an abelian $Z$-theory~\cite{Carrasco:2016ldy}. Our result for NLSM has less nonzero numerators while still retains the polynomial form in terms of Mandelstam variables. At the amplitude level, one may achieve this by employing a few BCJ relations, while it is still curious to explore the implication in string theory. For YM numerators, one can compare our result with some previous explicit constructions, for example, \cite{Mafra:2011kj,Mafra:2015vca}, and further explore the connection.

Last but not least, since our numerators are systematically derived from the spanning trees, one may ask what is the relation between our spanning trees and the more familiar Feynman diagrams. Actually, a deeper question to ask is instead of merely a technical tool, whether we can give some physical meaning to these spanning trees based on which we can reformulate the field theory. To gain more insight along this direction, further study in the loop level numerators will be very helpful.

\acknowledgments
We would like to thank Bo Feng for very enlightening discussions. YD would like to acknowledge National Natural Science Foundation of China under Grant Nos. 11105118, 111547310, as well as the 351 program of Wuhan University.
\appendix

%%%%%%%%%%%%%%%%%%%%%%%%%%%%%%%%%%%%%%%%%%%%%%%%%%%%%%%%%%%%%%%%%%%%%%%
\section{Laplace expansion of reduced Pfaffian}\label{sec:laplace}
%%%%%%%%%%%%%%%%%%%%%%%%%%%%%%%%%%%%%%%%%%%%%%%%%%%%%%%%%%%%%%%%%%%%%%%

We start with deleting the $1$-st and $n$-th row and column in $\Psi$ such that Eq.~\eqref{eq:Pfp} becomes:
\begin{equation}
\label{eq:Pfp1n}
    \Pfp(\Psi)\cong\frac{(-1)^{n}}{\sigma_{n1}}\Pf(\Psi_{1,n}^{1,n})\,.
\end{equation}
As we will see later, this choice leads to an expansion in terms of the KK basis with the position of particle $1$ and $n$ fixed. We emphasize that $\Pfp(\Psi)$ being independent of deleted rows and columns only holds after we impose the momentum conservation, transversality and scattering equation.

Next, we give the recursive relation for $\Pfp(\Psi)$ that finally leads to Eq.~\eqref{eq:expansion}. Suppose the set $\mathsf{A}$ and $\mathsf{H}=\{h_1\ldots h_m\}$ form a split of $\{2\ldots n-1\}$:
\begin{equation*}
    \mathsf{A}\cup\mathsf{H}=\{2\ldots n-1\}\,,
\end{equation*}
and $\pmb\alpha$ is a permutation of $\mathsf{A}$ (namely, $\pmb{\alpha}=\{\alpha(1),\alpha(2)\ldots \alpha(s)\}\in S_{\mathsf{A}}$), we can define the following quantity:
\begin{equation}
    \Theta\left(1,\pmb\alpha,n\,|\{h_1\ldots h_m\}\right)=\Pf\left(\begin{array}{cccc}
    A_{\mathsf{H}} & -(\pmb{v}_{\pmb\alpha})^{T} & -(C_{\mathsf{H}})^{T} & -(\pmb{c}_{n})^{T} \\
    \pmb{v}_{\pmb\alpha} & 0 & \pmb{u}_{\pmb\alpha} & w_{\pmb\alpha} \\
    C_{\mathsf{H}} & -(\pmb{u}_{\pmb\alpha})^{T} & B_{\mathsf{H}} & -(\pmb{b}_{n})^{T} \\
    \pmb{c}_{n} & -w_{\pmb{\alpha}} & \pmb{b}_{n} & 0 \\
    \end{array}\right)\,.
\end{equation}
In this expression, $A_{\mathsf{H}}$, $B_{\mathsf{H}}$ and $C_{\mathsf{H}}$ come from the EYM integrand \eqref{eq:EYMpf} with graviton set $\mathsf{H}$, and the vector $\pmb{c}_n$ and $\pmb{b}_n$ come from the last row of the matrix $C$ and $B$ given in \eqref{eq:ABC}:
\begin{align}
    (\pmb{c}_n)_{i}=\frac{\epsilon_n\cdot k_i}{\sigma_{ni}}& &(\pmb{b}_n)_{i}=\frac{\epsilon_n\cdot \epsilon_i}{\sigma_{ni}}& & (i\in\mathsf{H})\,.
\end{align}
More importantly, the vector $\pmb{v}_{\pmb\alpha}$ and $\pmb{u}_{\pmb\alpha}$ have the form:
\begin{align}
    (\pmb{v}_{\pmb\alpha})_i &=\left[\epsilon_{1}\cdot F_{\alpha(1)}\ldots F_{\alpha(s)}\cdot k_i\right]\PT(1,\pmb\alpha,i)\frac{\sigma_{i1}}{\sigma_{n1}}\nonumber\\
    (\pmb{u}_{\pmb\alpha})_i &=\left[\epsilon_{1}\cdot F_{\alpha(1)}\ldots F_{\alpha(s)}\cdot \epsilon_i\right]\PT(1,\pmb\alpha,i)\frac{\sigma_{i1}}{\sigma_{n1}}\,,
\end{align}
with $i\in\mathsf{H}$. Finally, the entry $w_{\pmb\alpha}$ has the form:
\begin{equation}
    w_{\pmb\alpha}=\left[\epsilon_1\cdot F_{\alpha(1)}\ldots F_{\alpha(s)}\cdot\epsilon_n\right]\PT\left(1,\pmb\alpha,n\right)\,.
\end{equation}
Once we expand $\Theta$ along the row of $\pmb{v}$ and $\pmb{u}$, we can derive a recursive relation:
\begin{align}
    \Theta\left(1,\pmb\alpha,n\,|\,\{h_1\ldots h_m\}\right)&=\left(-1\right)^{m}\left[\epsilon_1\cdot F_{\alpha(1)}\ldots F_{\alpha(s)}\cdot\epsilon_n\right]\PT\left(1,\pmb\alpha,n\right)\Pf(\Psi_{\mathsf{H}})\nonumber\\
    &\quad+\left(-1\right)^{m}\sum_{i=1}^{m}\Theta\left(1,\pmb\alpha,h_i,n\,|\,\{h_1\ldots\slashed{h}_{i}\ldots h_m\}\right)\,,
\end{align}
with $\mathsf{H}=\{h_1\ldots h_m\}$.This recursion starts when $\mathsf{A}$ is empty:
\begin{equation}
\label{eq:Pfprecursion}
    \frac{(-1)^{n}}{\sigma_{n1}}\Pf(\Psi_{1,n}^{1,n})=(-1)^{n}\Theta\left(1,n\,|\,\{2,3\ldots n-2\}\right)\,,
\end{equation}
and ends when $\mathsf{H}$ is empty:
\begin{align}
    \Theta\left(1,\pmb\alpha,n\,|\,\varnothing\right)=\left[\epsilon_1\cdot F_{\alpha(2)}\ldots F_{\alpha(n-2)}\cdot\epsilon_n\right]\PT\left(1,\pmb\alpha,n\right)& &\pmb\alpha\in S_{\{2\ldots n-2\}}\,.
\end{align}
We remark here that this recursive expansion holds at the off-shell level, namely, it is an algebraic identity with no requirement on $k$, $\epsilon$ and $\sigma_i$. In contrary, Eq.~\eqref{eq:Pfp1n} has physical sense only if we impose the on-shell conditions. Now it is straightforward to check that Eq.~\eqref{eq:Pfprecursion} leads to our proclaimed result Eq.~\eqref{eq:expansion}.

%%%%%%%%%%%%%%%%%%%%%%%%%%%%%%%%%%%%%%%%%%%%%%%%%%%%
\section{Deriving the YM graphic rules from the EYM expansion}\label{sec:derivation}
In~\cite{Teng:2017tbo}, a set of graphic rules are proposed to evaluate the coefficients $C_{\pmb\rho}(\pmb\sigma)$ in the expansion of EYM amplitudes in terms of YM ones. Using the notations in the current paper, we can write the expansion as:
\begin{equation}
\label{eq:EYMexpansion2}
    A_{s+2,m}^{\text{EYM}}(1,\pmb\alpha,n\,|\,\mathsf{H})=\sum_{\pmb\sigma\in S_{\mathsf{H}}}\sum_{\pmb\gamma\in\pmb\alpha\shuffle\pmb\sigma}C_{\pmb\rho}(\pmb\sigma)A_{n}^{\text{YM}}(1,\pmb\gamma,n)\,,
\end{equation}
where $\{1,\pmb\alpha,n\}$ is a set of $s+2$ color ordered gluons and $\mathsf{H}=\{h_1,h_2\ldots h_m\}$ is the set of gravitons. We first repeat briefly the graphic rules here:
\begin{itemize}
    \item We treat all the gluons as a single vertex $\mathsf{g}$ and draw the increasing trees with respect to the order:
    \begin{equation*}
        \mathsf{g}\prec h_{\sigma(1)}\prec\ldots\prec h_{\sigma(|\mathsf{H}|)}\,.
    \end{equation*}
    We denote the set of increasing tress as $IT(\pmb{\sigma})$.
    \item The evaluation of each tree is almost the same as {\bf Step 3} in Section~\ref{sec:rules}, except that when a path originates from the root $\mathsf{g}$, we close that chain by $Y_{h_i}(\pmb\gamma)$ at the graviton, say $h_i$, that is connected to the root. For each $\pmb\gamma\in\pmb\alpha\shuffle\pmb\sigma$, $Y_{h_i}(\pmb\gamma)$ is the sum of the gluon momenta coming ahead of $h_i$.
\end{itemize}
Consequently, if the expression associated with a tree $i\in IT(\pmb\sigma)$ is $T(i,\pmb\gamma)$, we can rewrite Eq.~\eqref{eq:EYMexpansion2} as:
\begin{equation}
    A_{s+2,m}^{\text{EYM}}(1,\pmb\alpha,n\,|\,\mathsf{H})=\sum_{\pmb\sigma\in S_{\mathsf{H}}}\sum_{\pmb\gamma\in\pmb\alpha\shuffle\pmb\sigma}\sum_{i\in IT(\pmb{\sigma})}T(i,\pmb\gamma)A_{n}^{\text{YM}}(1,\pmb\gamma,n)\,.
\end{equation}
We emphasize that the structure of $T(i,\pmb\gamma)$ only depends on $\pmb\sigma$, while  $\pmb\gamma$ only determines the values of $Y$'s. Now according to Eq.~\eqref{eq:BCJconstruction}, we need to fully expand the sum over permutations and shuffles in the above equation, in order to extract the contribution to a certain $\pmb\beta\in S_{\{2\ldots n-1\}}$. As a result, we will eliminate the $Y$'s in favor of the manifest gluon momenta. This can be achieved by the following arrangement.

Suppose we have a tree $T(i,\pmb\gamma)$ whose root $\mathsf{g}$ is connected to $r$ gravitons, we can rewrite this $T$ as:
\begin{align}
    \adjustbox{raise=-1.5cm}{\begin{tikzpicture}
        \filldraw [red,thick] (0,0) circle (2pt) node[below=1pt]{$1$} -- ++(1,0) circle (2pt) node[below=1pt]{$\alpha(1)$} -- ++(1,0) circle (2pt) node[below=1pt]{$\alpha(2)$} ++(0.5,0) node{$\cdots$} ++(0.5,0) circle (2pt) -- ++(1,0) circle (2pt) node[below=1pt]{$\alpha(s)$} -- ++(1,0) circle (2pt) node[below=1pt]{$n$};
        \draw [thick] (-0.5,-0.8) rectangle (4.5,0.8) node[below left=1pt]{$\mathsf{g}$};
        \filldraw [thick] (1,0.8) -- ++(0,1) circle (2pt) node [below left=1pt]{$h_1$};
        \draw [thick] (1,1.8) -- ++(-0.5,0.5) (1,1.8) -- ++(0.5,0.5);
        \filldraw [thick,fill=gray!30!white] (0.3,2.3) rectangle (1.7,3);
        \node at (1,2.65) {$B_1$};
        \filldraw [thick] (4,0.8) -- ++(0,1) circle (2pt) node [below left=1pt]{$h_r$};
        \draw [thick] (4,1.8) -- ++(-0.5,0.5) (4,1.8) -- ++(0.5,0.5);
        \filldraw [thick,fill=gray!30!white] (3.3,2.3) rectangle (4.7,3);
        \node at (4,2.65) {$B_r$};
        \node at (2.5,1.5) {$\cdots\cdots$};
    \end{tikzpicture}}\;=T(i,\pmb\gamma)=\prod_{s=1}^{r}\left[B_s\cdot Y_{h_s}(\pmb{\gamma})\right]\,.
\end{align}
In this way, we have singled out the $\pmb\gamma$ dependence in the trees. Now we can expand $Y$ for each $\pmb\gamma$. For simplicity, we only consider two branches. Starting with the branch $B_2$, the factor $(B_2\cdot k_{\alpha(j)})$ will appear for all orderings in
\begin{equation*}
    \big\{\alpha(1)\ldots\alpha(j),\{h_2B_2\}\shuffle\{\alpha(j+1)\ldots\alpha(s)\}\big\}\,.
\end{equation*}
Adding in $B_1$, then we have $(B_1\cdot k_{\alpha(j)})(B_2\cdot k_{\alpha(i)})$ for all orderings in
\begin{equation}
    \Big\{\alpha(1)\ldots\alpha(i),\{h_1B_1\}\shuffle\big\{\alpha(i+1)\ldots\alpha(j),\{h_2B_2\}\shuffle\{\alpha(j+1)\ldots\alpha(s)\}\big\}\Big\}\,.
\label{eq:2tree}
\end{equation}
Thus we assign $(B_1\cdot k_{\alpha(i)})(B_2\cdot k_{\alpha(j)})$ to the tree structure
\begin{equation}
\label{eq:2tvalue}
    \adjustbox{raise=-1.5cm}{\begin{tikzpicture}
        \filldraw [thick] (1,0) -- ++(0,1.8) circle (2pt) node [below left=1pt]{$h_1$};
        \draw [thick] (1,1.8) -- ++(-0.5,0.5) (1,1.8) -- ++(0.5,0.5);
        \filldraw [thick,fill=gray!30!white] (0.3,2.3) rectangle (1.7,3);
        \node at (1,2.65) {$B_1$};
        \filldraw [thick] (2,0) -- ++(1,1.8) circle (2pt) node [below right=1pt]{$h_2$};
        \draw [thick] (3,1.8) -- ++(-0.5,0.5) (3,1.8) -- ++(0.5,0.5);
        \filldraw [thick,fill=gray!30!white] (2.3,2.3) rectangle (3.7,3);
        \node at (3,2.65) {$B_2$};
        \filldraw [red,thick] (-1,0) circle (2pt) node[below=1pt]{$1$} -- (0,0) circle (2pt) node[below=1pt]{$\alpha(1)$} ++(0.5,0) node{$\cdots$} ++(0.5,0) circle (2pt) node[below=1pt]{$\alpha(i)$} ++(0.5,0) node{$\cdots$} ++(0.5,0) circle (2pt) node[below=1pt]{$\alpha(j)$} ++(0.5,0) node{$\cdots$} ++(0.5,0) circle (2pt) -- ++(1,0) circle (2pt) node[below=1pt]{$\alpha(s)$} -- ++(1,0) circle (2pt) node[below=1pt]{$n$};
    \end{tikzpicture}}\;=(B_1\cdot k_{\alpha(i)})(B_2\cdot k_{\alpha(j)})\,.
\end{equation}
If we start with a spanning tree without specifying gluons and gravitons, then the gluons can be simply identified as those on the path from the leaf $n$ to the root $1$. Thus for all $\pmb{\gamma}$ in Eq.~\eqref{eq:2tree}, the tree~\eqref{eq:2tvalue} is an increasing tree. On the other hand, the tree~\eqref{eq:2tvalue} contributes to all $\pmb\gamma$ in \eqref{eq:2tree}. The generalization to more branches are very straightforward: there are just more levels of shuffles. Actually, such a tree $\mathsf{T}$ will contribute to all those $\pmb\beta$ that $\mathsf{T}\in IT(\pmb\beta)$. The algorithm to work out these $\pmb\beta$'s are given in~\cite{Teng:2017tbo}.

To wrap up, this calculation derives the rules in {\bf Step 1} and {\bf Step 2} of Section~\ref{sec:rules} on how the increasing tree structure and the separation of gluons from gravitons emerge. It also shows how to close a graviton tree when it lands on a specific gluon, as stated in {\bf Step 3} of Section~\ref{sec:rules}.

%%%%%%%%%%%%%%%%%%%%%%%%%%%%%%%%%%%%%%%%%%%%%%%%%%%%
\section{Explicit five-point DDM form BCJ numerators for YM}\label{sec:numerator}
%%%%%%%%%%%%%%%%%%%%%%%%%%%%%%%%%%%%%%%%%%%%%%%%%%%%

In Section~\ref{sec:5p}, we have calculated $n^{\text{YM}}(1,\{234\},5)$. In this appendix, we list all the rest of the numerators:
\begingroup
\allowdisplaybreaks
\begin{align}
    n^{\text{YM}}(1,\{243\},5)&=-(\epsilon_1\cdot\epsilon_5)(\epsilon_4\cdot k_1)(\epsilon_3\cdot k_1)(\epsilon_2\cdot k_1)+(\epsilon_1\cdot F_4\cdot\epsilon_5)(\epsilon_3\cdot k_1)(\epsilon_2\cdot k_1)\nonumber\\*% row 1 (243)
    &\quad+(\epsilon_1\cdot F_3\cdot\epsilon_5)(\epsilon_4\cdot k_1)(\epsilon_2\cdot k_1)+(\epsilon_1\cdot F_2\cdot\epsilon_5)(\epsilon_4\cdot k_1)(\epsilon_3\cdot k_1)\nonumber\\ % row 1 (243)
    &\quad-(\epsilon_1\cdot\epsilon_5)(\epsilon_4\cdot F_2\cdot k_1)(\epsilon_3\cdot k_1)+(\epsilon_1\cdot F_3\cdot\epsilon_5)(\epsilon_4\cdot F_2\cdot k_1)\nonumber\\* % row 2 (243)
    &\quad-(\epsilon_1\cdot F_2\cdot F_4\cdot\epsilon_5)(\epsilon_3\cdot k_1)+(\epsilon_1\cdot F_2\cdot\epsilon_5)(\epsilon_4\cdot k_2)(\epsilon_3\cdot k_1)\nonumber\\ % row 2 (243)
    &\quad-(\epsilon_1\cdot\epsilon_5)(\epsilon_4\cdot k_1)(\epsilon_3\cdot k_4)(\epsilon_2\cdot k_1)-(\epsilon_1\cdot F_4\cdot F_3\cdot\epsilon_5)(\epsilon_2\cdot k_1)\nonumber\\* % row 3 (243)
    &\quad+(\epsilon_1\cdot F_4\cdot\epsilon_5)(\epsilon_3\cdot k_4)(\epsilon_2\cdot k_1)+(\epsilon_1\cdot F_2\cdot\epsilon_5)(\epsilon_4\cdot k_1)(\epsilon_3\cdot k_4)\nonumber\\ % row 3 (243)
    &\quad-(\epsilon_1\cdot\epsilon_5)(\epsilon_4\cdot F_2\cdot k_1)(\epsilon_3\cdot k_4)+(\epsilon_1\cdot F_2\cdot F_4\cdot F_3\cdot\epsilon_5)\nonumber\\* % row 4 (243)
    &\quad-(\epsilon_1\cdot F_2\cdot F_4\cdot\epsilon_5)(\epsilon_3\cdot k_4)+(\epsilon_1\cdot F_2\cdot\epsilon_5)(\epsilon_4\cdot k_2)(\epsilon_3\cdot k_4)\nonumber\\ % row 4 (243)
    &\quad-(\epsilon_1\cdot\epsilon_5)(\epsilon_3\cdot F_2\cdot k_1)(\epsilon_4\cdot k_1)-(\epsilon_1\cdot F_2\cdot F_3\cdot\epsilon_5)(\epsilon_4\cdot k_1)\nonumber\\* % row 5 (243)
    &\quad+(\epsilon_1\cdot F_4\cdot\epsilon_5)(\epsilon_3\cdot F_2\cdot k_1)+(\epsilon_1\cdot F_2\cdot\epsilon_5)(\epsilon_3\cdot k_2)(\epsilon_4\cdot k_1)\nonumber\\ % row 5 (243)
    &\quad-(\epsilon_1\cdot\epsilon_5)(\epsilon_4\cdot F_2\cdot k_1)(\epsilon_3\cdot k_2)-(\epsilon_1\cdot F_2\cdot F_3\cdot\epsilon_5)(\epsilon_4\cdot k_2)\nonumber
    \\* % row 6 (243)
    &\quad-(\epsilon_1\cdot F_2\cdot F_4\cdot\epsilon_5)(\epsilon_3\cdot k_2)+(\epsilon_1\cdot F_2\cdot\epsilon_5)(\epsilon_4\cdot k_2)(\epsilon_3\cdot k_2)\\ % row 6 (243)
%%%%%%%%%%%%%%%%%%%%%%%%%%%%%%%%%%%%%%%%%%%%%%%%%%%%%%%%%%%%%%%%%
    n^{\text{YM}}(1,\{324\},5)&=-(\epsilon_1\cdot\epsilon_5)(\epsilon_4\cdot k_1)(\epsilon_3\cdot k_1)(\epsilon_2\cdot k_1)+(\epsilon_1\cdot F_4\cdot\epsilon_5)(\epsilon_3\cdot k_1)(\epsilon_2\cdot k_1)\nonumber\\* % row 1 (324)
    &\quad+(\epsilon_1\cdot F_3\cdot\epsilon_5)(\epsilon_4\cdot k_1)(\epsilon_2\cdot k_1)+(\epsilon_1\cdot F_2\cdot\epsilon_5)(\epsilon_4\cdot k_1)(\epsilon_3\cdot k_1)\nonumber\\ % row 1 (324)
    &\quad-(\epsilon_1\cdot\epsilon_5)(\epsilon_4\cdot k_1)(\epsilon_3\cdot k_1)(\epsilon_2\cdot k_3)+(\epsilon_1\cdot F_4\cdot\epsilon_5)(\epsilon_3\cdot k_1)(\epsilon_2\cdot k_3)\nonumber\\* % row 2 (324)
    &\quad-(\epsilon_1\cdot F_3\cdot F_2\cdot\epsilon_5)(\epsilon_4\cdot k_1)+(\epsilon_1\cdot F_3\cdot\epsilon_5)(\epsilon_4\cdot k_1)(\epsilon_2\cdot k_3)\nonumber\\ % row 2 (324)
    &\quad-(\epsilon_1\cdot\epsilon_5)(\epsilon_4\cdot F_2\cdot k_1)(\epsilon_3\cdot k_1)-(\epsilon_1\cdot F_2\cdot F_4\cdot\epsilon_5)(\epsilon_3\cdot k_1)\nonumber\\* % row 3 (324)
    &\quad+(\epsilon_1\cdot F_2\cdot\epsilon_5)(\epsilon_4\cdot k_2)(\epsilon_3\cdot k_1)+(\epsilon_1\cdot F_3\cdot\epsilon_5)(\epsilon_4\cdot F_2\cdot k_1)\nonumber\\ % row 3 (324)
    &\quad-(\epsilon_1\cdot\epsilon_5)(\epsilon_4\cdot F_2\cdot F_3\cdot k_1)+(\epsilon_1\cdot F_3\cdot F_2\cdot F_4\cdot\epsilon_5)\nonumber\\* % row 4 (324)
    &\quad-(\epsilon_1\cdot F_3\cdot F_2\cdot\epsilon_5)(\epsilon_4\cdot k_2)+(\epsilon_1\cdot F_3\cdot\epsilon_5)(\epsilon_4\cdot F_2\cdot k_3)\nonumber\\ % row 4 (324)
    &\quad-(\epsilon_1\cdot\epsilon_5)(\epsilon_4\cdot F_3\cdot k_1)(\epsilon_2\cdot k_1)-(\epsilon_1\cdot F_3\cdot F_4\cdot\epsilon_5)(\epsilon_2\cdot k_1)\nonumber\\* % row 5 (324)
    &\quad+(\epsilon_1\cdot F_2\cdot\epsilon_5)(\epsilon_4\cdot F_3\cdot k_1)+(\epsilon_1\cdot F_3\cdot\epsilon_5)(\epsilon_4\cdot k_3)(\epsilon_2\cdot k_1)\nonumber\\ % row 5 (324)
    &\quad-(\epsilon_1\cdot\epsilon_5)(\epsilon_4\cdot F_3\cdot k_1)(\epsilon_2\cdot k_3)-(\epsilon_1\cdot F_3\cdot F_4\cdot\epsilon_5)(\epsilon_2\cdot k_3)\nonumber\\* % row 6 (324)
    &\quad-(\epsilon_1\cdot F_3\cdot F_2\cdot\epsilon_5)(\epsilon_4\cdot k_3)+(\epsilon_1\cdot F_3\cdot\epsilon_5)(\epsilon_4\cdot k_3)(\epsilon_2\cdot k_3)\\ % row 6 (324)
%%%%%%%%%%%%%%%%%%%%%%%%%%%%%%%%%%%%%%%%%%%%%%%%%%%%%%%%%%%%%%%%%%%%%%
    n^{\text{YM}}(1,\{342\},5)&=-(\epsilon_1\cdot\epsilon_5)(\epsilon_4\cdot k_1)(\epsilon_3\cdot k_1)(\epsilon_2\cdot k_1)+(\epsilon_1\cdot F_4\cdot\epsilon_5)(\epsilon_3\cdot k_1)(\epsilon_2\cdot k_1)\nonumber\\* % row 1 (342)
    &\quad+(\epsilon_1\cdot F_3\cdot\epsilon_5)(\epsilon_4\cdot k_1)(\epsilon_2\cdot k_1)+(\epsilon_1\cdot F_2\cdot\epsilon_5)(\epsilon_4\cdot k_1)(\epsilon_3\cdot k_1)\nonumber\\ % row 1 (342)
    &\quad-(\epsilon_1\cdot\epsilon_5)(\epsilon_4\cdot F_3\cdot k_1)(\epsilon_2\cdot k_1)+(\epsilon_1\cdot F_2\cdot\epsilon_5)(\epsilon_4\cdot F_3\cdot k_1)\nonumber\\* % row 2 (342)
    &\quad-(\epsilon_1\cdot F_3\cdot F_4\cdot\epsilon_5)(\epsilon_2\cdot k_1)+(\epsilon_1\cdot F_3\cdot\epsilon_5)(\epsilon_4\cdot k_3)(\epsilon_2\cdot k_1)\nonumber\\ % row 2 (342)
    &\quad-(\epsilon_1\cdot\epsilon_5)(\epsilon_4\cdot k_1)(\epsilon_3\cdot k_1)(\epsilon_2\cdot k_4)-(\epsilon_1\cdot F_4\cdot F_2\cdot\epsilon_5)(\epsilon_3\cdot k_1)\nonumber\\* % row 3 (342)
    &\quad+(\epsilon_1\cdot F_4\cdot\epsilon_5)(\epsilon_3\cdot k_1)(\epsilon_2\cdot k_4)+(\epsilon_1\cdot F_3\cdot\epsilon_5)(\epsilon_4\cdot k_1)(\epsilon_2\cdot k_4)\nonumber\\ % row 3 (342)
    &\quad-(\epsilon_1\cdot\epsilon_5)(\epsilon_4\cdot F_3\cdot k_1)(\epsilon_2\cdot k_4)+(\epsilon_1\cdot F_3\cdot F_4\cdot F_2\cdot\epsilon_5)\nonumber\\* % row 4 (342)
    &\quad-(\epsilon_1\cdot F_3\cdot F_4\cdot\epsilon_5)(\epsilon_2\cdot k_4)+(\epsilon_1\cdot F_3\cdot\epsilon_5)(\epsilon_4\cdot k_3)(\epsilon_2\cdot k_4)\nonumber\\* % row 4 (342)
    &\quad-(\epsilon_1\cdot\epsilon_5)(\epsilon_4\cdot k_1)(\epsilon_3\cdot k_1)(\epsilon_2\cdot k_3)-(\epsilon_1\cdot F_3\cdot F_2\cdot\epsilon_5)(\epsilon_4\cdot k_1)\nonumber\\* % row 5 (342)
    &\quad+(\epsilon_1\cdot F_4\cdot\epsilon_5)(\epsilon_3\cdot k_1)(\epsilon_2\cdot k_3)+(\epsilon_1\cdot F_3\cdot\epsilon_5)(\epsilon_4\cdot k_1)(\epsilon_2\cdot k_3)\nonumber\\ % row 5 (342)
    &\quad-(\epsilon_1\cdot\epsilon_5)(\epsilon_4\cdot F_3\cdot k_1)(\epsilon_2\cdot k_3)-(\epsilon_1\cdot F_3\cdot F_2\cdot\epsilon_5)(\epsilon_4\cdot k_3)\nonumber\\* % row 6 (342)
    &\quad-(\epsilon_1\cdot F_3\cdot F_4\cdot\epsilon_5)(\epsilon_2\cdot k_3)+(\epsilon_1\cdot F_3\cdot\epsilon_5)(\epsilon_4\cdot k_3)(\epsilon_2\cdot k_3)\\ % row 6 (342)
%%%%%%%%%%%%%%%%%%%%%%%%%%%%%%%%%%%%%%%%%%%%%%%%%%%%%%%%%%%%%%%%%%%%%%%%%
    n^{\text{YM}}(1,\{423\},5)&=-(\epsilon_1\cdot\epsilon_5)(\epsilon_4\cdot k_1)(\epsilon_3\cdot k_1)(\epsilon_2\cdot k_1)+(\epsilon_1\cdot F_4\cdot\epsilon_5)(\epsilon_3\cdot k_1)(\epsilon_2\cdot k_1)\nonumber\\* % row 1 (423)
    &\quad+(\epsilon_1\cdot F_3\cdot\epsilon_5)(\epsilon_4\cdot k_1)(\epsilon_2\cdot k_1)+(\epsilon_1\cdot F_2\cdot\epsilon_5)(\epsilon_4\cdot k_1)(\epsilon_3\cdot k_1)\nonumber\\ % row 1 (423)
    &\quad-(\epsilon_1\cdot\epsilon_5)(\epsilon_4\cdot k_1)(\epsilon_3\cdot k_1)(\epsilon_2\cdot k_4)+(\epsilon_1\cdot F_3\cdot\epsilon_5)(\epsilon_4\cdot k_1)(\epsilon_2\cdot k_4)\nonumber\\* % row 2 (423)
    &\quad-(\epsilon_1\cdot F_4\cdot F_2\cdot\epsilon_5)(\epsilon_3\cdot k_1)+(\epsilon_1\cdot F_4\cdot\epsilon_5)(\epsilon_3\cdot k_1)(\epsilon_2\cdot k_4)\nonumber\\ % row 2 (423)
    &\quad-(\epsilon_1\cdot\epsilon_5)(\epsilon_4\cdot k_1)(\epsilon_3\cdot F_2\cdot k_1)-(\epsilon_1\cdot F_2\cdot F_3\cdot\epsilon_5)(\epsilon_4\cdot k_1)\nonumber\\* % row 3 (423)
    &\quad+(\epsilon_1\cdot F_2\cdot\epsilon_5)(\epsilon_4\cdot k_1)(\epsilon_3\cdot k_2)+(\epsilon_1\cdot F_4\cdot\epsilon_5)(\epsilon_3\cdot F_2\cdot k_1)\nonumber\\ % row 3 (423)
    &\quad-(\epsilon_1\cdot\epsilon_5)(\epsilon_4\cdot k_1)(\epsilon_3\cdot F_2\cdot k_4)+(\epsilon_1\cdot F_4\cdot F_2\cdot F_3\cdot\epsilon_5)\nonumber\\* % row 4 (423)
    &\quad-(\epsilon_1\cdot F_4\cdot F_2\cdot\epsilon_5)(\epsilon_3\cdot k_2)+(\epsilon_1\cdot F_4\cdot\epsilon_5)(\epsilon_3\cdot F_2\cdot k_4)\nonumber\\ % row 4 (423)
    &\quad-(\epsilon_1\cdot\epsilon_5)(\epsilon_4\cdot k_1)(\epsilon_3\cdot k_4)(\epsilon_2\cdot k_1)-(\epsilon_1\cdot F_4\cdot F_3\cdot\epsilon_5)(\epsilon_2\cdot k_1)\nonumber\\* % row 5 (423)
    &\quad+(\epsilon_1\cdot F_2\cdot\epsilon_5)(\epsilon_4\cdot k_1)(\epsilon_3\cdot k_4)+(\epsilon_1\cdot F_4\cdot\epsilon_5)(\epsilon_3\cdot k_4)(\epsilon_2\cdot k_1)\nonumber\\ % row 5 (423)
    &\quad-(\epsilon_1\cdot\epsilon_5)(\epsilon_4\cdot k_1)(\epsilon_3\cdot k_4)(\epsilon_2\cdot k_4)-(\epsilon_1\cdot F_4\cdot F_3\cdot\epsilon_5)(\epsilon_2\cdot k_4)\nonumber\\* % row 6 (423)
    &\quad-(\epsilon_1\cdot F_4\cdot F_2\cdot\epsilon_5)(\epsilon_3\cdot k_4)+(\epsilon_1\cdot F_4\cdot\epsilon_5)(\epsilon_3\cdot k_4)(\epsilon_2\cdot k_4)\\ % row 6 (423)
%%%%%%%%%%%%%%%%%%%%%%%%%%%%%%%%%%%%%%%%%%%%%%%%%%%%%%%%%%%%%%%%%%%%%%%%%%%%
    n^{\text{YM}}(1,\{432\},5)&=-(\epsilon_1\cdot\epsilon_5)(\epsilon_4\cdot k_1)(\epsilon_3\cdot k_1)(\epsilon_2\cdot k_1)+(\epsilon_1\cdot F_4\cdot\epsilon_5)(\epsilon_3\cdot k_1)(\epsilon_2\cdot k_1)\nonumber\\* % row 1 (432)
    &\quad+(\epsilon_1\cdot F_3\cdot\epsilon_5)(\epsilon_4\cdot k_1)(\epsilon_2\cdot k_1)+(\epsilon_1\cdot F_2\cdot\epsilon_5)(\epsilon_4\cdot k_1)(\epsilon_3\cdot k_1)\nonumber\\ % row 1 (432)
    &\quad-(\epsilon_1\cdot\epsilon_5)(\epsilon_4\cdot k_1)(\epsilon_3\cdot k_4)(\epsilon_2\cdot k_1)+(\epsilon_1\cdot F_2\cdot\epsilon_5)(\epsilon_4\cdot k_1)(\epsilon_3\cdot k_4)\nonumber\\* % row 2 (432)
    &\quad-(\epsilon_1\cdot F_4\cdot F_3\cdot\epsilon_5)(\epsilon_2\cdot k_1)+(\epsilon_1\cdot F_4\cdot\epsilon_5)(\epsilon_3\cdot k_4)(\epsilon_2\cdot k_1)\nonumber\\ % row 2 (432)
    &\quad-(\epsilon_1\cdot\epsilon_5)(\epsilon_4\cdot k_1)(\epsilon_3\cdot k_1)(\epsilon_2\cdot k_3)-(\epsilon_1\cdot F_3\cdot F_2\cdot\epsilon_5)(\epsilon_4\cdot k_1)\nonumber\\* % row 3 (432)
    &\quad+(\epsilon_1\cdot F_3\cdot\epsilon_5)(\epsilon_4\cdot k_1)(\epsilon_2\cdot k_3)+(\epsilon_1\cdot F_4\cdot\epsilon_5)(\epsilon_3\cdot k_1)(\epsilon_2\cdot k_3)\nonumber\\ % row 3 (432)
    &\quad-(\epsilon_1\cdot\epsilon_5)(\epsilon_4\cdot k_1)(\epsilon_3\cdot k_4)(\epsilon_2\cdot k_3)+(\epsilon_1\cdot F_4\cdot F_3\cdot F_2\cdot\epsilon_5)\nonumber\\* % row 4 (432)
    &\quad-(\epsilon_1\cdot F_4\cdot F_3\cdot\epsilon_5)(\epsilon_2\cdot k_3)+(\epsilon_1\cdot F_4\cdot\epsilon_5)(\epsilon_3\cdot k_4)(\epsilon_2\cdot k_3)\nonumber\\% rwo 4 (432)
    &\quad-(\epsilon_1\cdot\epsilon_5)(\epsilon_4\cdot k_1)(\epsilon_3\cdot k_1)(\epsilon_2\cdot k_4)-(\epsilon_1\cdot F_4\cdot F_2\cdot\epsilon_5)(\epsilon_3\cdot k_1)\nonumber\\* % row 5 (432)
    &\quad+(\epsilon_1\cdot F_3\cdot\epsilon_5)(\epsilon_4\cdot k_1)(\epsilon_2\cdot k_4)+(\epsilon_1\cdot F_4\cdot\epsilon_5)(\epsilon_3\cdot k_1)(\epsilon_2\cdot k_4)\nonumber\\ % row 5 (432)
    &\quad-(\epsilon_1\cdot\epsilon_5)(\epsilon_4\cdot k_1)(\epsilon_3\cdot k_4)(\epsilon_2\cdot k_4)-(\epsilon_1\cdot F_4\cdot F_2\cdot\epsilon_5)(\epsilon_3\cdot k_4)\nonumber\\* % row 6 (432)
    &\quad-(\epsilon_1\cdot F_4\cdot F_3\cdot\epsilon_5)(\epsilon_2\cdot k_4)+(\epsilon_1\cdot F_4\cdot\epsilon_5)(\epsilon_3\cdot k_4)(\epsilon_2\cdot k_4)\,. % row 6 (432)
\end{align}
\endgroup

%%%%%%%%%%%%%%%%%%%%%%%%%%%%%%%%%%%%%%%%%%%%%%%%%%%%%%%%%%%%%%%%%
\section{Explicit six-point DDM form BCJ numerators for NLSM}\label{sec:NLSMnum}
%%%%%%%%%%%%%%%%%%%%%%%%%%%%%%%%%%%%%%%%%%%%%%%%%%%%%%%%%%%%%%%%%

In Section~\ref{sec:NLSMexample}, we calculated $n(1,\{2345\},6)$ for NLSM. Now we list all the rest $23$ DDM form numerators evaluated with the reference order $\pmb\rho=\{5432\}$, which are calculated in the same way:
\begingroup
\allowdisplaybreaks
\begin{align}
    n^{\text{NLSM}}(1,\{2354\},6)&=(k_5\cdot k_3)(k_3\cdot k_1)(k_4\cdot k_2)(k_2\cdot k_1)+(k_5\cdot k_2)(k_2\cdot k_1)(k_4\cdot k_3)(k_3\cdot k_1)\nonumber\\* % 2354
    &\quad+(k_5\cdot k_2)(k_2\cdot k_1)(k_4\cdot k_3)(k_3\cdot k_2)\\ % 2354
    n^{\text{NLSM}}(1,\{2435\},6)&=(k_5\cdot k_3)(k_3\cdot k_1)(k_4\cdot k_2)(k_2\cdot k_1)+(k_5\cdot k_3)(k_3\cdot k_4)(k_4\cdot k_2)(k_2\cdot k_1)\nonumber\\* % 2435
    &\quad+(k_5\cdot k_4)(k_4\cdot k_1)(k_3\cdot k_2)(k_2\cdot k_1) \\ % 2435
    n^{\text{NLSM}}(1,\{2453\},6)&=(k_5\cdot k_4)(k_4\cdot k_1)(k_3\cdot k_2)(k_2\cdot k_1)\\ % 2453
    n^{\text{NLSM}}(1,\{2534\},6)&=(k_5\cdot k_2)(k_2\cdot k_1)(k_4\cdot k_3)(k_3\cdot k_1)+(k_5\cdot k_2)(k_2\cdot k_1)(k_4\cdot k_3)(k_3\cdot k_5)\nonumber\\* % 2534
    &\quad+(k_5\cdot k_2)(k_2\cdot k_1)(k_4\cdot k_3)(k_3\cdot k_2) \\ % 2534
    n^{\text{NLSM}}(1,\{2543\},6)&=0 \\ % 2543
    n^{\text{NLSM}}(1,\{3245\},6)&=(k_5\cdot k_3)(k_3\cdot k_1)(k_4\cdot k_2)(k_2\cdot k_1)+(k_5\cdot k_4)(k_4\cdot k_2)(k_2\cdot k_3)(k_3\cdot k_1)\nonumber\\* % 3245
    &\quad+(k_5\cdot k_3)(k_3\cdot k_1)(k_4\cdot k_2)(k_2\cdot k_3)+(k_5\cdot k_2)(k_2\cdot k_1)(k_4\cdot k_3)(k_3\cdot k_1) \\ % 3245
    n^{\text{NLSM}}(1,\{3254\},6)&=(k_5\cdot k_2)(k_2\cdot k_1)(k_4\cdot k_3)(k_3\cdot k_1)+(k_5\cdot k_3)(k_3\cdot k_1)(k_4\cdot k_2)(k_2\cdot k_1)\nonumber\\* % 3254
    &\quad+(k_5\cdot k_3)(k_3\cdot k_1)(k_4\cdot k_2)(k_2\cdot k_3) \\ % 3254
    n^{\text{NLSM}}(1,\{3425\},6)&=(k_5\cdot k_2)(k_2\cdot k_1)(k_4\cdot k_3)(k_3\cdot k_1)+(k_5\cdot k_2)(k_2\cdot k_4)(k_4\cdot k_3)(k_3\cdot k_1) \\ % 3425
    n^{\text{NLSM}}(1,\{3452\},6)&=0 \\ % 3452
    n^{\text{NLSM}}(1,\{3524\},6)&=(k_5\cdot k_3)(k_3\cdot k_1)(k_4\cdot k_2)(k_2\cdot k_1)+(k_5\cdot k_3)(k_3\cdot k_1)(k_4\cdot k_2)(k_2\cdot k_5)\nonumber\\* % 3524
    &\quad+(k_5\cdot k_3)(k_3\cdot k_1)(k_4\cdot k_2)(k_2\cdot k_3) \\ % 3524
    n^{\text{NLSM}}(1,\{3542\},6)&=0 \\ % 3542
    n^{\text{NLSM}}(1,\{4235\},6)&=(k_5\cdot k_4)(k_4\cdot k_1)(k_3\cdot k_2)(k_2\cdot k_1)+(k_5\cdot k_3)(k_3\cdot k_2)(k_2\cdot k_4)(k_4\cdot k_1)\nonumber\\* % 4235
    &\quad+(k_5\cdot k_4)(k_4\cdot k_1)(k_3\cdot k_2)(k_2\cdot k_4) \\ % 4235
    n^{\text{NLSM}}(1,\{4253\},6)&=(k_5\cdot k_4)(k_4\cdot k_1)(k_3\cdot k_2)(k_2\cdot k_1)+(k_5\cdot k_4)(k_4\cdot k_1)(k_3\cdot k_2)(k_2\cdot k_4) \\ % 4253
    n^{\text{NLSM}}(1,\{4325\},6)&=(k_5\cdot k_2)(k_2\cdot k_3)(k_3\cdot k_4)(k_4\cdot k_1) \\ % 4325
    n^{\text{NLSM}}(1,\{4352\},6)&=0 \\ % 4352
    n^{\text{NLSM}}(1,\{4523\},6)&=(k_5\cdot k_4)(k_4\cdot k_1)(k_3\cdot k_2)(k_2\cdot k_1)+(k_5\cdot k_4)(k_4\cdot k_1)(k_3\cdot k_2)(k_2\cdot k_5)\nonumber\\* % 4523
    &\quad+(k_5\cdot k_4)(k_4\cdot k_1)(k_3\cdot k_2)(k_2\cdot k_4)\\ % 4523
    n^{\text{NLSM}}(1,\{4532\},6)&=0 \\ % 4532
    n^{\text{NLSM}}(1,\{5234\},6)&=n^{\text{NLSM}}(1,\{5243\},6)=n^{\text{NLSM}}(1,\{5324\},6)=0 \\ % 5
    n^{\text{NLSM}}(1,\{5342\},6)&=n^{\text{NLSM}}(1,\{5423\},6)=n^{\text{NLSM}}(1,\{5432\},6)=0
\end{align}
\endgroup

% BIBLIOGRAPHY
% use BIBTEX if you want
\bibliographystyle{JHEP}
\bibliography{Refs}

\end{document}